\documentclass[leqno,authoryear,1p]{article}
\usepackage[utf8]{inputenc}
\usepackage{times}
\usepackage{enumitem}
\usepackage{bbold}
\usepackage{tikz-cd}
\usepackage{tikz}
\usetikzlibrary{decorations.pathmorphing}
\usetikzlibrary{quotes}
\usepackage{dsfont}
\usepackage{comment}
\usepackage{mathrsfs}
\usepackage{slashed}
\usepackage[french,german,english]{babel}
\usepackage{soul}
\usepackage{amssymb,amsmath}
\usepackage{natbib}
\title{Logic of gauge}
\author{Alexander Afriat}
\begin{document}
\maketitle
\begin{abstract}
\noindent The logic of gauge theory is considered by tracing its development from general relativity to Yang-Mills theory, through Weyl's two gauge theories. A handful of elements---which for want of better terms can be called \emph{geometrical justice}, \emph{matter wave}, \emph{second clock effect}, \emph{twice too many energy levels}---are enough to produce Weyl's second theory; and from there, all that's needed to reach the Yang-Mills formalism is a \emph{non-Abelian structure group} (say $\mathbb{SU}\textrm{(}N\textrm{)}$).
\end{abstract}

\begin{quote}\begin{quote}
\tableofcontents
\end{quote}\end{quote}
\thispagestyle{empty}
\clearpage
\setcounter{page}{1}

\section{Introduction}
This article is an attempt to answer the question(s) ``How did we get gauge theory? What's the logic of the historical process that produced such a theory?'' In a sentence, my answer will be something like:
\begin{quote}
Once general relativity was rectified by \emph{geometrical justice}, a (necessarily relativistic) \emph{matter wave} was enough---bearing in mind the \emph{second clock effect}, and the \emph{too many energy levels} of Dirac's theory---to produce Weyl's second theory, which, with a \emph{non-Abelian structure group}, leads to the Yang-Mills formalism.
\end{quote}
That's a gist, largely unintelligible for the time being. Other answers are of course possible too.

It makes sense to begin with general relativity, and I have chosen not to go beyond Yang-Mills theory, concentrating most of my attention on the transition from Weyl's first gauge theory \textsf{\small{W18}} to his second \textsf{\small{W29}}. Surprisingly little is needed, \emph{in a purely logical sense}, to go all the way from \textsf{\small{GR}} to \textsf{\small{YM}}---which is not to diminish all the impressive scientific creativity involved in between. What I propose is a necessarily \emph{a posteriori}\footnote{I cannot help writing in 2017---not before at any rate. The article will be read no earlier than 2017, by readers whose habits, mental categories and means of understanding were formed in recent decades, not in the nineteenth century. Any attempt to understand the ideas of 1929 will have to be made using the cognitive resources available to \emph{actual} or \emph{possible}  readers, not to fancied readers bred in a fictitious past; so-called anachronisms can be as necessary as they are dangerous (the slope is indeed slippery, there's no denying that).} logical clarification or reconstruction---which may look more anachronistic than it really is---of the historical steps involved, without suggestions of condescending retrospective `trivialization': the point isn't that now, with the benefit of hindsight, we can see that it was all trivial, or even inevitable, borne along by inexorable historical necessity; but that with the benefit of hindsight a certain logically simplified or even `sanitized' representation of the development can be given, which sheds light on the full evolution, in all its bewildering detail and archaic colour. My purpose is not to replace a close perusal of the texts with misleading anachronisms; but to supplement their study with a clarified representation of the evolution they express, in terms more intelligible to certain modern readers. The usefulness of a map for jungle exploration does not preclude jungle exploration---it can hardly oblige its reader to stay at home with the map.

The facts, the texts, underdetermine the logical representation, which cannot be unique; the reconstruction I have chosen is intended to capture the basic structure of the evolution as cleanly as possible.

``Logic'' and cognates have been used in many different ways---\emph{John} 1:1, \citet{Hegel}, \citet{Popper} \emph{etc}. The meaning I have in mind, which should emerge in the sequel, has to do with `derivation' (of a theory for instance), and `what elements are needed to derive.'

``History'' and cognates have also been used in many different ways; there are many ways to treat, understand, study, describe, approach the past. It is a sociological fact that different communities behave and communicate in different ways. I have left ``history'' out of the title to avoid confusing a community that uses the word in its own very special manner.

There may be a tradeoff between ``clear and distinct,'' `Cartesian' intelligibility and a kind of historical fidelity. If there is, this paper can be seen as an attempted optimization of sorts, subject to the operative constraints.\footnote{\emph{Best of both worlds} is more the idea than \emph{having one's cake and eating it}---which suggests an attempt to violate inescapable constraints.} An extreme example illustrating what I mean by ``Cartesian intelligibility'' might be a good (but perhaps anachronistic) `historical' synopsis in a recent text by a differential geometer. At the other end one can struggle with and try to render, with comprehensive zeal, all the ambiguities, irrelevancies, idiosyncrasies and misunderstandings of the original texts. Such a struggle can provide \emph{one special kind} of historical understanding; but other kinds or aspects, none the less legitimate, may be left out. Of course the problem can be eliminated, with discreet but trenchant\footnote{An appropriate image might be the discreet amputation of a mischievous or perhaps embarrassing limb.} expeditiousness, by carefully adjusting the scope of \emph{history} to leave out awkward desiderata: ``that's not history (any more)!''

The point is not to satisfy all conceivable desiderata, but to avoid the questionable neglect of certain reasonable ends. An appropriate form of Cartesian intelligibility seems a goal worth at least \emph{pursuing} (\emph{success} being quite another matter).

The importance of \emph{explication} in historical analysis should not be underestimated; and it is best to explicate in terms intelligible to the (necessarily modern) reader. Questions like ``what on earth is meant here?''~often arise, and deserve answers, which should be \emph{less} obscure than the original. Weyl, for instance, can be extremely murky; should that murkiness be faithfully handed on to the reader, to avoid anachronism, in a spirit of historical fidelity?

Needless to say, an account that's already too tidy (again, ``anachronistic'' is the word that's used in these cases) for many historians won't be sanitized enough for many mathematical physicists who use the terms \emph{Weyl structure}, \emph{Weyl tensor etc}.\ every day---but that's a matter of tribal behaviour and the sociology of scientific communities. `Treading a fine line' isn't even the right sociological metaphor; far from being fine, it is of considerable---only \emph{negative}---width.

One anachronism is the symbol $A$ I use for the length connection (and electromagnetic potential). Weyl creates \textsf{\small{W18}} for one purpose: the anholonomic propagation of length. So the whole point of the length connection is its curvature, it cannot be an exact differential $df$. If Weyl's notation $df$ were taken literally, his theory would be pointless, there would be no electricity, Einstein's objection would be vacuous and senseless \emph{etc}. To denote an object that cannot be (necessarily) exact it therefore seems best to avoid a notation expressing exactness. Another anachronism is my use of a \emph{single} letter $\mathbb{W}$ to denote Weyl's `material' group, for which he uses many letters, in fact several words.\footnote{\citet{Weyl1929} p.~333: ``\foreignlanguage{german}{man beschränke sich auf solche lineare Transformationen $U$ von $\psi_1$, $\psi_2$, deren Determinante den absoluten Betrag $1$ hat.}''\label{Wgroup}} Notation is bound to be problematic here: too close an adherence to Weyl's would produce misunderstandings, whereas entirely modern notation, however intelligible today, would just be too remote. I have attempted a compromise (with obvious perils): without losing touch with Weyl I have tried to remain intelligible to modern readers.

Many anachronisms are deplorable. But there is something wrong with the common inference \emph{some anachronisms are to be avoided therefore anachronisms are all to be avoided}. Telling good and bad anachronisms apart can admittedly be hard, it is easiest to proscribe them altogether (baby with bathwater)\hspace{2pt}\ldots\hspace{1pt}spares the trouble of distinguishing~\dots

Since the transition (\S\ref{GeomJust}) from general relativity to Weyl's first gauge theory (1918, \S\ref{W18}) has been amply discussed in \citet{Pais1982}, \citet{Vizgin}, \citet{Scholz1994, Scholz1995, Scholz2001, Scholz2004, Scholz2011b}, \citet{Cao}, \citet{Hawkins}, \citet{ColemanK}, \citet{Sigurdsson2001}, \citet{Ryckman2003, RyckmanBC, RyckmanReign, Ryckman2009}, \citet{PenroseRoad} and \citet{Afriat}, I concentrate on the next step, which took Weyl to his second theory (\S\ref{W29})---mainly determined by the new undulatory ontology (\S\ref{NUO}) introduced by \citet{deBroglie}, \citet{Dirac}, \citet{Schroedinger} and others. Mainly but not wholly: Weyl had every reason to keep the electricity, gravity and \emph{abstract} gauge structure of his first theory; but now with three ingredients (matter, electricity, gravity), three different gauge relations were possible, two of which (electricity-matter, electricity-gravity) were more plausible than the third (matter-gravity). Einstein's objection (1918, \S\ref{clock2}), the \emph{second clock effect}, ruled out the old gauge relation ((\ref{gauge})\&(\ref{conforme})) between electricity and gravity, leaving the new relation ((\ref{gauge})\&(\ref{phase})) between electricity and matter. Weyl's objection that four-component theory provided \emph{twice too many energy levels} (\S\ref{DWT}) is only relevant to his own story, of how he reached his two-component theory of 1929, and not to \textsf{\small{YM}} (\S\ref{YangMills})---which by no means favours Weyl's two-component theory over Dirac's with four.

\textsf{\small{YM}} is more relevant here as abstract \emph{mathematical} physics (like symplectic geometry with differential forms) than as genuine theoretical \emph{physics}. The story may indeed appear to assume a somewhat unexpected---perhaps even fictitious---character in \S\ref{YangMills}. Since the numerous physical details that ultimately did produce \textsf{\small{YM}} seem rather foreign to the spirit and purpose of my account, I have chosen to deal only with the `purely logical' transition from \textsf{\small{W29}} to \textsf{\small{YM}}.

\section{Weyl's first gauge theory\label{W18}}
\subsection{Geometrical justice\label{GeomJust}}
Again, it makes sense to start with \textsf{\small{GR}}.\footnote{\citet{Einstein1916}} \citet{Levi-Civita} saw that the connection determined by Einstein's covariant derivative transported the \emph{direction} of a vector anholonomically, but not its \emph{length}, which was left unchanged.\footnote{See \citet{RyckmanBC} p.~80, \citet{Ryckman2009} p.~288.} This was unfair, protested Weyl---length deserved the same treatment as direction.\footnote{See \citet{Afriat} for details of this `geometrical justice'---which can also be understood in terms of group extensions (see \citet{Scholz2004} pp.~183, 189, 191-2, \citet{Scholz2011a} pp.~195, \citet{Scholz2011b}, third page of the paper): since a Levi-Civita connection subjects direction to $\mathbb{SO}^+\textrm{(}1,3\textrm{)}$ but length to the (group containing only the) identity $\mathds{1}$, it is only fair to extend the identity by the dilations, yielding $\mathds{1}\times\mathbb{R}=\mathbb{R}$---which (unlike $\mathds{1}$) allows length anholonomies and therefore geometrical justice. The total group, for direction and length together, is the extension $\mathbb{SO}^+\textrm{(}1,3\textrm{)}\times\mathbb{R}$ giving the relativistic similarities. But the group $\mathbb{W}$ Weyl uses in 1929 is (globally) not the extension $\mathbb{SL}\textrm{(}2,\mathbb{C}\textrm{)}\times\mathbb{U}\textrm{(}1\textrm{)}$; see \S\ref{anachronism}. \citet{Ryckman2003, RyckmanBC, RyckmanReign, Ryckman2009} provides an alternative account of Weyl's agenda.\label{extensions}} To remedy he proposed a more general theory that propagated length just as anholonomically as direction. The resulting \emph{congruent} transport would also be governed by a connection, which Weyl defined quite generally as a bilinear mapping between neighbouring points: linear in the thing propagated and in the direction of propagation. A connection transporting the (squared) length $l$ from $a=\gamma\textrm{(0)}$ to its neighbour\footnote{Which is so close to $a$ it practically belongs to the tangent space $T_aM$; see \citet{PdMuN} p.~28, \citet{Weyl1931} p.~52.} $b=\gamma\textrm{(1)}$ along the world-line\footnote{$\mathfrak{I}\subset\mathbb{R}$ is an appropriate interval containing $0$ and $1$.} $\gamma:\mathfrak{I}\rightarrow M$, $\tau\mapsto \gamma\textrm{(}\tau\textrm{)}$ would therefore be a real-valued\footnote{Here the structure group is the multiplicative group $\mathbb{R}$ of dilations, generated by the Lie algebra 
$\langle\mathbb{R},+,\textrm{[}\hspace{1pt}\cdot\hspace{1pt},\cdot\hspace{1pt}\textrm{]}\rangle$ or rather $\langle\mathbb{R},+\rangle$; the Lie product $\textrm{[}\hspace{1pt}\cdot\hspace{1pt},\cdot\hspace{1pt}\textrm{]}$ vanishes since real numbers commute.} one-form\footnote{Einstein's summation convention will sometimes be used.} $A=A_\mu dx^\mu$ applied to the tangent vector $\dot{\gamma}=\dot{\gamma}^\mu\partial_\mu\in T_aM$ and multiplied by the initial length $l_a$, yielding the increment\footnote{I often use angular brackets $\langle\alpha,X\rangle$ to denote the value of the form or covector $\alpha$ at the vector $X$. Bras and kets (which presuppose an appropriate natural pairing) will also be useful, especially where inner $\langle\eta|\zeta\rangle$ and outer $|\zeta\rangle\langle\eta |$ products both arise.}
$$\delta l=l_b-l_a=l_a\langle A,\dot{\gamma}\rangle=l_a A_{\mu}\dot{\gamma}^{\mu}$$
added to $l_a$.\footnote{\emph{Cf}.\ \citet{Ryckman2009} pp.~290-1.} The final length $l_b$ is $l_a(1+\langle A,\dot{\gamma}\rangle)$---unless $a$ and $b$ are too far apart for $\gamma$ to remain straight in between, in which case $l_b$ is
$$
l_a\exp\int_\gamma A.
$$
Congruent transport can also be expressed by the differential equation $\partial_\tau l=\langle A,\dot{\gamma}\rangle l$, where $\partial_\tau$ differentiates along $\dot{\gamma}$.

To deal with the geometrical injustice that $A$ was to remedy, the curvature
\begin{equation}\label{curvature}
F=\frac{1}{2}F_{\mu\nu}dx^\mu\wedge dx^\nu =dA=\frac{1}{2}(\partial_\mu A_\nu-\partial_\nu A_\mu)dx^\mu\wedge dx^\nu
\end{equation}
 cannot (necessarily) vanish---unlike the three-form
\begin{equation}\label{homogeneous}
dF=d^2A=\frac{1}{6}(\partial_\mu F_{\nu\sigma}+\partial_\nu F_{\sigma\mu}+\partial_\sigma F_{\mu\nu})\hspace{1pt}dx^\mu\wedge dx^\nu\wedge dx^\sigma\textrm{,}
\end{equation}
which does. Seeing all this, Weyl couldn't help thinking\footnote{See \citet{Eddington} p.~175, \citet{Scholz2001} p.~75, \citet{Ryckman2003} p.~92, \citet{RyckmanReign} p.~158.} of the electromagnetic four-potential $A$, the Faraday two-form $F=dA$ and Maxwell's two homogeneous equations\footnote{$\nabla\cdot\mathbf{B}=0$ and $\nabla\times\mathbf{E}+\partial_t\mathbf{B}=0$} $dF=0$: he had unified gravity and electromagnetism---by mistake!\footnote{\citet{RyckmanBC} p.~61: ``[\,\dots] Weyl did not start out with the objective of unifying gravitation and electromagnetism, but sought to remedy a perceived blemish in Riemannian `infinitesimal' geometry. The resulting `unification' was, as it were, serendipitous.'' See also p.~63, \citet{Ryckman2003} p.~86, \citet{RyckmanReign} pp.~149-54, 158, \citet{Ryckman2009} pp.~287-94.} And indeed Einstein would soon point out the mistake: the anholonomy on which Weyl had built his theory is not found in nature, as we shall see in \S\ref{clock2}.

\subsection{Gauge}\label{GaugeSection}
\noindent Weyl sought, then, to rectify general relativity by means of the curvature (\ref{curvature}), which ensured geometrical justice: vanishing \emph{Streckenkrümmung} (length curvature) $F$ led to \emph{holonomic} congruent transport which, alongside \emph{anholonomic} parallel transport, was manifestly unfair.

Differentiation is destructive, or rather irreversible; what $d$ destroys (through its kernel) is the freedom
\begin{equation}\label{gauge}
A\mapsto A'=A+d\lambda
\end{equation}
invisible to $F=dA=dA'$, in the sense that the inverse image $[A]=d^{-1}F$ of $F$ under $d$ is the whole equivalence class given by the equivalence relation $A\sim\textrm{(}A+d\lambda\textrm{)}$, where the function $\lambda$ assigns a \emph{single} (and hence path-independent) value to each point of $M$. If $A$ only served to produce the curvature $F$, with no other role, (\ref{gauge}) would be vacuous; but $A$ appears elsewhere too, notably in the law of propagation
\begin{equation}\label{2}
\nabla g=A\otimes g\textrm{,}
\end{equation}
which is not indifferent to (\ref{gauge}), $g$ being the metric. To make (\ref{2}) invariant, (\ref{gauge}) therefore has to be balanced by
\begin{equation}\label{conforme}
g\mapsto g'=e^{\lambda}g\emph{,}
\end{equation}
whose origin is thus accounted for.\footnote{\emph{Cf}.\ \citet{Weyl1931} p.~54: ``\foreignlanguage{german}{insbesondere konnte ich nichts a priori Einleuchtendes vorbringen zugunsten der Koppelung des willkürlichen additiven Gliedes ${\partial\lambda}/{\partial x_p}$, das nach der Erfahrung in den Komponenten des elektromagnetischen Potentials steckt, mit dem von der klassischen Geometrie geforderten Eichfaktor $e^{\lambda}$.}''} The point of Weyl's theory is not the `holonomic'\footnote{But \citet{Dirac1931} gives infinitesimal (and indeed globally path-dependent, anholonomic) meaning to the similar expression $e^{i\beta}$ in his equation (3), where $\beta$ cannot be a \emph{single}-valued function on space-time.} transformation (\ref{conforme}), which should, despite much emphasis in the literature,\footnote{The misunderstandings go at least as far back as \citet[first published in 1920]{Eddington}, who first, at the top of p.~169, goes out of his way to explain the \emph{an}holonomic propagation of length; which then, just a few lines on, gets propagated in the very way that was to be avoided, subject to the very restriction that was to be overcome: ``a definite unit of interval, or gauge, \emph{at every point of space and time}. [\,\dots] when the comparison depends on the route taken, exact equality is not definable; and we have therefore to admit that the \emph{exact} standards are laid down at every point independently.'' \emph{Cf}.\ \citet{Dirac1931} p.~63: ``We may assume that $\gamma$ has no definite value at a particular point, but only a definite difference in values for any two points. We may go further and assume that this difference is not definite unless the two points are neighbouring. For two distant points there will then be a definite phase difference only relative to some curve joining them and different curves will in general give different phase differences.'' Indeed there are three cases, not two: [1] no variation at all, [2] holonomic variation, [3] anholonomic variation. Eddington has rightly understood that Weyl wants to go beyond [1]. But that leaves the other \emph{two}---the whole point of \textsf{\scriptsize{W18}} being that Weyl wants to go beyond [2] as well. If one's bent on conflation, the \emph{first} two cases can be more or less conflated in \textsf{\scriptsize{W18}}: $F$ vanishes ([1], [2]), or not ([3])---\emph{how} it vanishes is hardly the point. Eddington seems to feel that conflation is needed somewhere, and duly conflates the \emph{last} two instead. The tradition he may or may not have founded has had considerable and perhaps growing success---it persists to this day, with an ample, zealous following, and no sign of abating; \emph{cf}.\ \citet{WGA}, especially footnotes 5 and 9, about an equivalent misunderstanding.} by no means be taken as a premiss, being more of a consequence than a postulate. The real premiss is geometrical justice---direction \emph{and length} both deserve anholonomic transport. Weyl did define conformally invariant objects, maybe quite a few, but (\ref{2}) is not among them, nor indeed is \textsf{\small{W18}}.

To see how the metric can correct (only locally) the dilations generated by the connection, we can take a unit length $l_a=g_a(V_a,V_a)=1$
at point $a\in M$; and a connection which annihilates the tangent vector $\dot{\gamma}$ directed towards $b$ nearby, so that $\langle A,\dot{\gamma}\rangle$ vanishes. As the connection produces no increment $\delta l$ in this case, the length $l_b=g_b(V_b,V_b)=1$ remains unaffected and nothing need be done to the metric $g_b$ at $b$.

But the increment $\delta l$ needn't vanish; in general (\ref{conforme}) is required to maintain the same numerical value $1=l_b$, by adapting the yardstick at $b$. Any exact term $d\lambda$ added to the connection will be balanced by a conformal factor $e^{\lambda}$, and \emph{vice versa}; the (holonomic) recalibrations (\ref{conforme}) of the metric only correct the dilations produced by $d\lambda$ (and not the anholonomic dilations due to a curved connection).

To spell out the implications of \emph{geometrical justice} in \textsf{\small{W18}} one can write: \emph{geometrical justice} $\Rrightarrow$ anholonomic connection $A\Rrightarrow$ curvature $F=dA\Rrightarrow$ indifference of $F$ to (\ref{gauge})\,$;$ sensitivity of (\ref{2}) to (\ref{gauge}) $\Rrightarrow$ indifference of (\ref{2}) to \{(\ref{gauge}) balanced by (\ref{conforme})\}.
Such compensation is typical of a gauge theory: an invariant expression (here (\ref{2})) is sensitive to a first transformation, and to a second as well---but indifferent to the two together, if their variations are appropriately constrained, and balance one another.\footnote{See \citet{Ryckman2003} p.~77.}

So far, then, we have two primitive logical elements
\begin{enumerate}
\item \textsf{\small{GR}}: \emph{general relativity}
\item \textsf{\small{GJ}}: \emph{geometrical justice}
\end{enumerate}
which together yield Weyl's theory of electricity and gravity:
$$
\textsf{\small{GR \& GJ}}\Rrightarrow\textsf{\small{W18}}.
$$
The next will be \textsf{\small{MW}}:~\emph{matter wave} and \textsf{\small{SC}}:~(avoid) \emph{second clock effect}. The latter is essentially Einstein's objection, which should now be considered.

\subsection{Einstein's objection\label{clock2}}
\noindent The tangent of a world-line's \emph{image} $\bar{\gamma}\subset M$ only has a direction, it is a full ray; the length
$$l=\|\dot{\gamma}\|^2=g\textrm{(}\dot{\gamma},\dot{\gamma}\textrm{)}$$
of the tangent vector $\dot{\gamma}={d\gamma}/{d\tau}$ is given by the parameter rate ${\partial\gamma}/{\partial\tau}$. If the values of the parameter are identified with the readings of a clock describing $\gamma$, the length $l$ giving the proper ticking rate should remain constant---the hands of a good clock don't accelerate. But far from remaining constant, lengths in Weyl's theory aren't even integrable:
$$l_b\textrm{(}\gamma\textrm{)}=l_a\exp\int_\gamma A$$
depends on $\gamma$---whereas an exact connection $A=d\mu$ would give
$$l_b=l_a\exp{\int_a^bd\mu}=l_a\exp\Delta\mu$$
along any path, $\Delta\mu$ being the difference $\mu\textrm{(}b\textrm{)}-\mu\textrm{(}a\textrm{)}$ between the final and initial values of $\mu$. In addition to the \emph{first} clock effect (Langevin's twins) already present in Einstein's theory, Weyl's theory therefore involves a \emph{second} clock effect expressed in the anholonomy of ticking rates.

Einstein objected that \emph{nature provides integrable clocks}.\footnote{Letter to Weyl dated 15 April 1918: ``\foreignlanguage{german}{So schön Ihre Gedanke ist, muss ich doch offen sagen, dass es nach meiner Ansicht ausgeschlossen ist, dass die Theorie der Natur entspricht. Das $ds$ selbst hat nämlich reale Bedeutung. Denken Sie sich zwei Uhren, die relativ zueinander ruhend neben einander gleich rasch gehen. Werden sie voneinander getrennt, in beliebiger Weise bewegt und dann wieder zusammen gebracht, so werden sie wieder gleich (rasch) gehen, d. h. ihr relativer Gang hängt nicht von der Vorgeschichte ab. Denke ich mir zwei Punkte $P_1$ \& $P_2$ die durch eine zeitartige Linie verbunden werden können. Die an $P_1$ \& $P_2$ anliegenden zeitartigen Elemente $ds_1$ und $ds_2$ können dann durch mehrere zeitartigen Linien verbunden werden, auf denen sie liegen. Auf diesen laufende Uhren werden ein Verhältnis $ds_1:ds_2$ liefern, welches von der Wahl der verbindenden Kurven unabhängig ist.---Lässt man den Zusammenhang des $ds$ mit Massstab- und Uhr-Messungen fallen, so verliert die Rel.\ Theorie überhaupt ihre empirische Basis}.'' Another letter to Weyl, four days later: ``\foreignlanguage{german}{wenn die Länge eines Einheitsmassstabes (bezw. die Gang-Geschwindigkeit einer Einheitsuhr) von der Vorgeschichte abhingen. Wäre dies in der Natur wirklich so, dann könnte es nicht chemische Elemente mit Spektrallinien von bestimmter Frequenz geben, sondern es müsste die relative Frequenz zweier (räumlich benachbarter) Atome der gleichen Art im Allgemeinen verschieden sein. Da dies nicht der Fall ist, scheint mir die Grundhypothese der Theorie leider nicht annehmbar, deren Tiefe und Kühnheit aber jeden Leser mit Bewunderung erfüllen muss}.''} Two clocks trace out a loop $\bar{\gamma}=\partial\omega$ enclosing a region $\omega$ (without holes): starting from the same point $a$ they describe worldlines $\gamma_1$, $\gamma_2$ that meet at $b$. They tick at the same rate if $A$ is exact, for then
$$\oint_{\partial\omega}d\mu=\iint_{\omega}d^2\mu$$
vanishes---since no holes are enclosed it is enough for $A$ to be closed,
 $$\oint_{\partial\omega}A=\iint_{\omega}dA$$
vanishes too provided $dA$ does. But if the loop encloses an electromagnetic field $F=dA$, one of the clocks will tick faster than the other once they're compared at $b$. In any case the theory didn't work: it rested from the outset on an anholonomy not seen in nature.\footnote{\emph{Cf}.\ \citet{Eddington} p.~175, \citet{Ryckman2009} p.~295.}

\section{Weyl's second gauge theory\label{W29}}
\noindent The setback of 1918, Einstein's objection (his preaching!\footnote{Letter to Seelig---quoted in \citet{Seelig} p.~274---in which Weyl quotes Einstein: ``\foreignlanguage{german}{So -- das heisst auf so spekulative Weise, ohne ein leitendes, anschauliches physikalisches Prinzip -- macht man keine Physik!}''})  put an end to Weyl's geometrical fantasies, to his `wilder days' as it were, producing a serious new empirical sobriety: ``All these geometrical leaps-in-the-air [\textsf{\small{W18}}] were premature, we return [\textsf{\small{W29}}] to the solid ground of physical facts.''\footnote{\citet{Weyl1931} p.~56: ``\foreignlanguage{german}{Alle diese geometrischen Luftsprünge waren verfrüht, wir kehren zurück auf den festen Boden der physikalischen Tatsachen.}'' \emph{Cf}.\ \citet{Scholz2011a} pp.~190-1.} Impressed at the unexpected transformation, Pauli speaks of ``revenge''\footnote{``Rache''; \citet{Pauli} p.~518: ``\foreignlanguage{german}{Als Sie früher die Theorie mit $g'_{ik}=\lambda g_{ik}$ machten, war dies reine Mathematik und unphysikalisch. Einstein konnte mit Recht kritisieren und schimpfen. Nun ist die Stunde der Rache für Sie gekommen; jetzt hat Einstein den Bock des Fernparallelismus geschossen, der auch nur reine Mathematik ist und nichts mit Physik zu tun hat, und Sie können schimpfen!}''}; with the zeal of a convert eager to trumpet his new convictions Weyl even insists that his new theory came \emph{straight out of experience}\footnote{\citet{Weyl1929} p.~331: ``\foreignlanguage{german}{Es scheint mir darum dieses nicht aus der Spekulation, sondern aus der Erfahrung stammende neue Prinzip der Eichinvarianz [\,\dots].}'' \citet{Weyl1931} p.~57: ``\foreignlanguage{german}{Das neue Prinzip ist aus der \emph{Erfahrung} erwachsen und resümiert einen gewaltigen, aus der Spektroskopie entsprungenen Erfahrungsschatz.}'' On Weyl's `empirical turn' see \citet{Scholz2004} pp.~165, 183, 191-3.} (and not out of his own hypercreative brain), \emph{directly derived from spectrographic data}\footnote{\citet{Weyl1931} p.~57: ``\foreignlanguage{german}{Dieses Transformationsgesetz der $\psi$ ist zuerst von \textsc{Pauli} aufgestellt worden und folgt mit unfehlbarer Sicherheit aus den spektroskopischen Tatsachen, genauer aus den Termdubletts der Alkalispektren und der Tatsache, daß die Dublettkomponenten nach Ausweis ihres Zeemaneffekts \emph{halbganze} innere Quantenzahlen besitzen}.''}  \dots

For his new theory takes account of the electron's spin---which in fact got there through the Dirac equation; and in Dirac's argument (1928) spin does not come straight out of experience\footnote{On the logical priority of relativity over spin \emph{cf}.\ \citet{GTQM} p.~193: ``\foreignlanguage{german}{Da die Möglichkeit einer solchen relativitätsinvarianten Gleichung für ein skalares $\psi$ nicht vorhanden ist, erscheint \emph{der Spin als ein durch die Relativitätstheorie notwendig gefordertes Phänomen}.}''} but out of a mathematical, æsthetic, \emph{a priori} principle, in much the same spirit as the geometrical justice that produced Weyl's first gauge theory.
\subsection{Matter\label{NUO}}
\noindent But let us go back a few years. As mentioned in the Introduction, \citet{deBroglie}, \citet{Dirac}, \citet{Schroedinger} \emph{et al}.\ had meanwhile produced an \emph{undulatory} world. Weyl had no reason to get rid of electricity or gravitation; to those existing ingredients he therefore had to add a matter wave, to update his ontology. As long as there was only gravity and electricity, the gauge relation (\ref{gauge})\&(\ref{conforme}) had to hold between \emph{them}; but now, with a third element, as many compensations were in principle possible, of which only two were plausible: the old relation between gravity and electricity, and a new one between electricity and matter. With (\ref{gauge})\&(\ref{conforme}) the theory would have remained subject to Einstein's objection---which the presence of the electron's mass $m$ (giving the wavelength\footnote{But here Planck's constant $h$ and the speed of light $c$---and even charge---are set equal to one.} ${h}/{mc}$ and frequency ${mc^2}/{h}$) in the Dirac equation made even more convincing,\footnote{\citet{Rice} p.~284: ``By this new situation, which introduces an atomic radius into the field equations themselves---but not until this step---my principle of \emph{gauge-invariance}, with which I had hoped to relate gravitation and electricity, is robbed of its support.'' \citet{Weyl1931} p.~55: ``\foreignlanguage{german}{Die Atomistik gibt uns ja absolute Einheiten für alle Maßgrößen an die Hand. [\,\dots] So geht in die \textsc{Dirac}sche Feldgesetze des Elektrons die "`Wellenlänge des Elektrons"', die Zahl ${h}/{mc}$, als eine absolute Konstante ein. Damit fällt das Grundprinzip meiner Theorie, das Prinzip von der Relativität der Längenmessung, dem Atomismus zum Opfer und verliert seine Überzeugungskraft.}'' See also \citet{Penrose} pp.~55-6.} by providing an absolute standard of length and time allowing the distant comparisons Weyl sought to rule out in \textsf{\small{W18}}.\footnote{See also \citet{Rice} p.~290.} The other possibility was left: (\ref{gauge}) with a quantum version of (\ref{conforme}),\footnote{\citet{Weyl1929} p.~331, \citet{Rice} p.~284: ``this principle has an equivalent in the quantum-theoretical field equations which is exactly like it in formal respects; the laws are invariant under the simultaneous replacement of $\psi$ by $e^{i\lambda}\psi$, $\varphi_\alpha$ by $\varphi_\alpha-{\partial\lambda}/{\partial x_\alpha}$ where $\lambda$ is an arbitrary real function of position and time.''} of which the most obvious\footnote{The transformation (\ref{phase}) is invisible `with respect to position,' or rather with respect to an observable compatible with the unitary operator determined by (\ref{phase}); \emph{cf}.\ \citet{GTQM1} p.~87. The requirement $\|\psi'\|=\|\psi\|$ is very natural but too weak to determine (\ref{phase}), being satisfied by \emph{any} unitary operator---not just those compatible with the representation (position, momentum or other) in which the wavefunction happens to be written.} was
\begin{equation}\label{phase}
\psi\mapsto \psi'=e^{i\lambda}\psi\textrm{,}
\end{equation}
where $\mathbb{U}\textrm{(}1\textrm{)}$ replaced the multiplicative group $\mathbb{R}$ of (\ref{conforme}).\footnote{\citet{Weyl1931} p.~55: ``\foreignlanguage{german}{In dem theoretischen Weltbild bedeutet die Verwandlung von $f_p$ in $-f_p$ eine objektive Änderung des metrischen Feldes; denn es ist etwas anderes, ob sich eine Strecke bei kongruenter Verpflanzung längs einer geschlossenen Bahn vergrößert oder verkleinert. Nach dem angenommenen Wirkungsgesetz aber ist die Entscheidung über das Vorzeichen der $f_p$ auf Grund der beobachteten Erscheinungen unmöglich. Hier enthält darum, in Widerstreit mit einem oben ausgesprochenen erkenntnistheoretischen Grundsatz, das theoretische Weltbild eine Verschiedenheit, welche sich auf keine Weise für die Wahrnehmung aufbrechen läßt.}'' P.~57: ``\foreignlanguage{german}{Die an der alten Theorie gerügte Unsicherheit des Vorzeichens $\pm f_p$ löst sich dadurch in das unbestimmte Vorzeichen der $\sqrt{-1}$ auf. Schon damals, als ich die alte Theorie aufstellte, hatte ich das Gefühl, daß der Eichfaktor die Form $e^{i\lambda}$ haben sollte; nur konnte ich dafür natürlich keine geometrische Deutung finden. Arbeiten von \textsc{Schrödinger} und \textsc{F.~London} stützten die Forderung durch die allmählich sich immer deutlicher abzeichnende Beziehung zur Quantentheorie.}'' See also \citet{GTQM} p.~89. \citet{Scholz2004} p.~193 associates the `geometry to matter' transition from (\ref{gauge})\&(\ref{conforme}) to (\ref{gauge})\&(\ref{phase}) with a transition from the \emph{a priori} fantasies of 1918 to the sober empiricism of 1929.} As $\psi$ was now part of a four-dimensional space-time theory, it could no longer obey the Schrödinger equation, which violates relativity by treating space and time very differently.\footnote{\citet{GTQM} pp.~187-8: ``\foreignlanguage{german}{Es ist klar, daß man zu einer befriedigenden Theorie des Elektrons nur kommen wird, wenn es gelingt, das Grundgesetz seiner Bewegung in der von der Relativitätstheorie geforderten, gegenüber Lorentz-Transformationen invarianten Form zu fassen.}''} Weyl adopted what amounted to a Dirac equation,\footnote{See \citet{Scholz2006} p.~470.} but cut in half: deprived of mass and the associated interweaving ($\sigma\upsilon\mu\pi\lambda\acute{\varepsilon}\kappa\varepsilon\iota\nu$, to use Weyl's term\footnote{\citet{ClassicalGroups} p.~165}) of component pairs.

We now have four logical elements:
\begin{enumerate}
\item \textsf{\small{GR}}: \emph{general relativity}
\item \textsf{\small{GJ}}: \emph{geometrical justice}
\item \textsf{\small{MW}}: \emph{matter wave}
\item \textsf{\small{SC}}: \emph{second clock effect};
\end{enumerate}
\textsf{\small{W29}} $\Lleftarrow$ \textsf{\small{W18 \& MW \& SC \&}} ?

A final element, \textsf{\small{EL}}: \emph{twice too many energy levels}, will be enough to produce \textsf{\small{W29}}, its basic structures at any rate.

\subsection{Spinors\label{DWT}}
\noindent The Hamiltonian $H$ of a free classical particle (of mass one-half) is the square $\|p\|^2=\langle p,\dot{q}\rangle$ of its momentum $p$; we can loosely write $H=p^2$. Momentum in quantum mechanics\footnote{See \citet{GTQM} pp.~45-6, 89.} is represented by differentiation $-i\nabla=\hat{p}$, so that the Hamiltonian $\hat{H}$ of a free quantum particle is $\hat{p}^2=-\nabla^2=-\nabla\hspace{-1pt}\cdot\hspace{-1pt}\nabla$. Writing $\hat{H}=-\nabla^2$ in $i\nabla_t\psi=\hat{H}\psi$ we obtain $i\nabla_t\psi=-\nabla^2\psi$, which shows that Schrödinger's equation violates relativity by differentiating space twice as much as time. But by what should it be replaced? The Klein-Gordon equation\footnote{\citet{GTQM} p.~186 attributes it to Louis de Broglie.} $(\square-m^2)\psi=0$ with d'Alembertian
$$\square =\partial_0^2-\partial_1^2-\partial_2^2-\partial_3^2$$
treats space about the same way as time, they have the right transformation properties; but $\square$ is `squared' and there are reasons to prefer a wave operator and especially a time derivative\footnote{\citet{GTQM} p.~188: ``\foreignlanguage{german}{Sie ist nicht im Einklang mit dem allgemeinen Schema der Quantenmechanik, welches verlangt, daß die zeitliche Ableitung nur in der ersten Ordnung auftritt}.'' P.~193: ``Legt man die de Brogliesche Wellengleichung für das skalare $\psi$ zugrunde, in welche die elektromagnetischen Potentiale $[A_\mu]$ durch die Regel [(\ref{covariant})] eingeführt sind, so ergibt sich aber für die elektrische Dichte ein Ausdruck, der außer $\psi$ die zeitliche Ableitung ${\partial \psi}/{\partial t}$ enthält und nichts mit der Ortswahrscheinlichkeit zu tun hat; sein Integral ist überhaupt keine Einzelform. Dies ist nach \emph{Dirac} das entscheidendste Argument dafür, daß die Differentialgleichungen des in einem elektromagnetischen Feld sich bewegenden Elektrons von 1.\ Ordnung in bezug auf die zeitliche Ableitung sein müssen.''} that aren't. In seeking a square root $\sqrt{\square}$ Dirac found $\slashed\partial=\gamma^\mu\partial_\mu$,
where the $\gamma^\mu$'s have the algebraic properties needed to get rid of the cross terms that appear when squaring. He therefore proposed the \emph{Dirac equation}\footnote{\citet{Dirac1928}}
\begin{equation}\label{Dirac}
(m-i\slashed\partial)\psi=0
\end{equation}
which not only treats the three spatial derivatives $\gamma^k\partial_k$ the same way as the time derivative $\gamma^0\partial_0$, but differentiates with respect to time only once.\footnote{\citet{GTQM} p.~190: ``\foreignlanguage{german}{Nach dem allgemeinen Schema der Quantenmechanik sollte, wie schon erwähnt, die Differentialgleichung für $\psi$ von 1.\ Ordnung hinsichtlich der zeitlichen Ableitung von $\psi$ sein. Gemäß dem Relativitätsprinzip kann sie aber dann auch nur die 1.\ Ableitungen nach den räumlichen Koordinaten enthalten.}''} The $\gamma^\mu$'s, which do not commute, cannot be numbers; they admit for instance the canonical representations
\begin{equation}\label{gamma}
\gamma^0\leftrightarrow\left(\hspace{-4pt}
    \begin{array}{cc}
      0 & \sigma^0 \\
      -\sigma^0 & 0 \\
    \end{array}
 \hspace{-4pt} \right)\qquad
\gamma^k\leftrightarrow\left(\hspace{-4pt}
    \begin{array}{cc}
      0 & \sigma^k \\
      \sigma^k & 0 \\
    \end{array}
 \hspace{-4pt} \right)\textrm{,}
\end{equation}
where all four quaternions $\sigma^\mu:\mathbb{C}^2\rightarrow\mathbb{C}^2$ are hermitian and unitary; $\sigma^0$ is the identity $\mathds{1}_2$, and the three traceless operators $\sigma^k$ satisfy $2i\sigma^j=\varepsilon_{jkl}[\sigma^k,\sigma^l]$.

The wave $\psi$ on which the $\gamma^\mu$'s act therefore has four (complex) components---\emph{embarras de richesses} which Weyl found most troubling\hspace{2pt}: ``\foreignlanguage{german}{doppelt zu viel Energie\-niveaus}''! The anti-diagonality of the $\gamma^\mu$'s governs the embarrassing excess by swapping the two two-spinors making up $\psi$. As the embarrassment is due to the \emph{sign} that distinguishes between the different interweavings\footnote{Symplectic for time, in the rather standard representation (\ref{gamma}), but simply `\texttt{NOT}' for space. The interweaving produced by four purely \texttt{NOT} $\gamma^{\mu}$'s (anti-diagonality with no minuses) would be pointless; Dirac's unusual hyperbolicity has to be expressed by one or more appropriately placed minuses: if the three spatial gammas have merely \texttt{NOT} anti-diagonality, $\gamma^0$ will be symplectically anti-diagonal.} produced by the $\gamma^\mu$'s, Weyl deals with it by choosing the only mass---none at all---that doesn't distinguish between plus and minus.\footnote{\citet{Weyl1929} p.~330-1: ``\foreignlanguage{german}{Die \so{Dirac}sche Theorie, in welcher das Wellenfeld des Elektrons durch ein Potential $\psi$ mit vier Komponenten beschrieben wird, gibt doppelt zu viel Energieniveaus; man sollte darum, ohne die relativistische Invarianz preiszugeben, zu den zwei Komponenten der \so{Pauli}schen Theorie zurückkehren können. Daran hindert das die Masse $m$ des Elektrons [\,\dots].}'' \citet{Rice} p.~292: ``The [mass] term (5) of the Dirac theory is, however, more doubtful. It must be admitted that if we retain it we can obtain all details of the line spectrum of the hydrogen atom---of one electron moving in the electrostatic field of a nucleus---in accord with what is known from experiment. But we obtain twice too much; if we replace the electron by a particle of the same mass and positive charge $+e$ (which admittedly does not exist in nature) the Dirac theory gives, contrary to all reason and experience, the same energy terms as for a negative electron, except for a change in sign. Obviously an essential change is here necessary.'' P.~294: ``Be bold enough to leave the term involving mass entirely out of the field equations.''} Without mass and half the components, (\ref{Dirac}) becomes
\begin{equation}\label{WEquation}
\sigma^\mu\partial_\mu\psi=0.
\end{equation}

We now have five logical elements:
\begin{enumerate}
\item \textsf{\small{GR}}: \emph{general relativity}
\item \textsf{\small{GJ}}: \emph{geometrical justice}
\item \textsf{\small{MW}}: \emph{matter wave}
\item \textsf{\small{SC}}: \emph{second clock effect}
\item \textsf{\small{EL}}: \emph{twice too many energy levels};
\end{enumerate}
\textsf{\small{W18 \& MW \& SC \& EL} $\Rrightarrow$ \small{W29}}.\footnote{\textsf{\scriptsize{W18 \& MW \& SC}} give something like Dirac-Maxwell theory in curved space-time.} The foundations having been laid, the rest will fall into place.

\subsection{Tetrads\label{tetrads}}
\noindent
The new arrival, \emph{matter}, is doubly unsettling: not only does it produce \emph{structural} upheaval, leading to a new gauge relation (now between matter and electricity); but the spinors used to represent matter even make Weyl alter the (mathematical) nature of gravity, now represented by tetrads---\emph{Achsenkreuze}---and no longer by the metric.

In \textsf{\small{W29}} we therefore see the appearance of a new material, quantum, complex world alongside the old gravitational, space-time, real world (leaving aside electricity for the time being, which indeed will presently emerge from their relationship). The two worlds are by no means independent, they are well entangled: since matter has to take account of the gravitational curvature of the space-time on which its spinors are defined, the material connection $\mathfrak{M}$ relies on the gravitational connection $\mathit{\Gamma}$---without, however, being exactly the same, as we shall soon see. A differential law like (\ref{WEquation}) compares values at neighbouring points; a background notion of `constancy' is needed for such a comparison to make sense; so a geometrically sensical field equation for spinors has to take account of space-time curvature.

Gravity in \textsf{\small{W18}} was represented by the metric $g=g_{\mu\nu}dx^{\mu}\otimes dx^{\nu}$, the oblique bases $dx^{\mu}$ being subject to $\mathbb{GL}\textrm{(}4,\mathbb{R}\textrm{)}$, which is much larger than $\mathbb{SO}^+\textrm{(}1,3\textrm{)}$ and has nothing to do with the group $\mathbb{SL}\textrm{(}2,\mathbb{C}\textrm{)}$ often used for relativistic spinors.\footnote{Weyl's spinors will in fact be subject to a slightly larger group but we can think of $\mathbb{SL}\textrm{(}2,\mathbb{C}\textrm{)}$ for the time being. Relevant group theory will be looked at more closely in \S\ref{electricity}.} To apply the same laws (or almost) to matter and gravity, Weyl replaces $\mathbb{GL}\textrm{(}4,\mathbb{R}\textrm{)}$ by the subgroup $\mathbb{SO}^+\textrm{(}1,3\textrm{)}$ locally isomorphic to $\mathbb{SL}\textrm{(}2,\mathbb{C}\textrm{)}$.\footnote{\citet{Rice} p.~285: ``The tensor calculus is not the proper mathematical instrument to use in translating the quantum-theoretic equations of the electron over into the \emph{general theory of relativity}. Vectors and terms are so constituted that the law which defines the transformation of their components from one Cartesian set of axes to another can be extended to the most general linear transformation, to an affine set of axes. That is not the case for the quantity $\psi$, however; this kind of quantity belongs to a representation of the rotation group which cannot be extended to the affine group. Consequently we cannot introduce components of $\psi$ relative to an arbitrary coordinate system in general relativity as we can for the electromagnetic potential and field strengths. We must rather describe the metric at a point $P$ by local Cartesian axes $e\textrm{(}\alpha\textrm{)}$ instead of by the $g_{pq}$. The wave field has definite components $\psi_1^+$, $\psi_2^+$; $\psi_1^-$, $\psi_2^-$ [full Dirac theory] relative to such axes, and we know how they transform on transition to any other Cartesian axes in $P$. The laws shall naturally be invariant under arbitrary rotation of the axes in $P$, and the axes at different points can be rotated independently of each other; they are in no way bound together.''} Why tetrads, then? Once $\mathbb{SO}^+\textrm{(}1,3\textrm{)}$ is chosen as the gravitational group, one might as well deal with entities it preserves. Of course $\mathbb{SO}^+\textrm{(}1,3\textrm{)}$ would preserve the geometrical relations of oblique bases too; but the orthonormality characterized by ones and noughts is distinguished. It would be pointless to adopt simple geometrical relations only to upset them with a large, disruptive group; but with a group that preserves geometrical simplicity, one might as well adopt that simplicity (orthonormality) from the start. Summing up: relativity~\textsf{\small{\& MW}} $\Rrightarrow$ something like the Dirac equation $\Rrightarrow$ relativistic spinors $\Rrightarrow$ something like $\mathbb{SL}\textrm{(}2,\mathbb{C}\textrm{)}\Rrightarrow$ space-time bases invariant under $\mathbb{SO}^+\textrm{(}1,3\textrm{)}\Rrightarrow$ orthonormal tetrads.

We'll now see how Weyl extracts electricity from the relationship between matter and gravity.

\subsection{Electricity\label{electricity}}
\subsubsection{Preliminary anachronisms\label{anachronism}}
Some geometry\footnote{See \citet{Funktionentheorie} pp.~7-15, \citet{GTQM} pp.~128-33, \citet{Smirnov} pp.~298-309, \citet{PenroseRindler} pp.~9-67, \citet{Needham} pp.~122-80.} is needed to understand what \citet{Weyl1929} is up to on pages 332-4.

Hermitian operators on $\mathbb{C}^2$ constitute a four-dimensional real vector space with scalar product $\langle\mathfrak{x},\mathfrak{y}\rangle=\frac{1}{2}\mathrm{Tr}\hspace{1pt}\textrm{(}\mathfrak{x}\hspace{1pt}\mathfrak{y}\textrm{)}$. The quaternions $\sigma^\mu$ we saw in (\ref{gamma}) make up a convenient orthonormal basis, $\langle\sigma^{\mu},\sigma^{\nu}\rangle=\delta^{\mu\nu}$. The real numbers $\mathfrak{x}^\mu=\langle\sigma^\mu,\mathfrak{x}\rangle$ can be viewed as the components of a space-time four-vector---whose hyperbolic squared length $\|\mathfrak{x}\|_{\eta}^2=\eta_{\mu\nu}\mathfrak{x}^\mu \mathfrak{x}^\nu$ is given by $\mathrm{det}\hspace{1pt}\mathfrak{x}$. The Lorentz group is defined by the isometric condition
$$
\mathbb{O}\textrm{(}1,3\textrm{)}=\{\Lambda\in\mathbb{GL}\textrm{(}4,\mathbb{R}\textrm{)}:\|\Lambda \mathfrak{x}\|_{\eta}^2=\|\mathfrak{x}\|_{\eta}^2\}\textrm{,}
$$
where the components of $\Lambda\mathfrak{x}$ are $\Lambda^\mu_\nu\mathfrak{x}^\nu$. An element $U$ of $\mathbb{GL}\textrm{(}2,\mathbb{C}\textrm{)}$, acting on a Hermitian operator $\mathfrak{x}$ as $\mathfrak{x}\mapsto U\mathfrak{x}\hspace{1pt}U^\dagger$,
preserves the determinant of $\mathfrak{x}$ if $\mathrm{det}\hspace{1pt}U=e^{i\theta}$, for then $\mathrm{det}\hspace{1pt}U\mathrm{det}(U^{\dagger})=1$, where the dagger denotes the adjoint. So the squared length $\|\mathfrak{x}\|_{\eta}^2=\mathrm{det}\hspace{1pt}\mathfrak{x}$, whose preservation characterizes the Lorentz group, is also preserved by the group\footnote{\citet{Weyl1929} p.~333: ``\foreignlanguage{german}{$U$ bewirkt an den $x_{\alpha}$ eine \so{Lorentztransformation}, d.\ i.\ eine reelle homogene lineare Transformation, welche die form $-x^2_0+x^2_1+x^2_2+x^2_3$ in sich überführt.}''}
$$
\mathbb{W}=\{U\in\mathbb{GL}\textrm{(}2,\mathbb{C}\textrm{)}: |\hspace{1pt}\mathrm{det}\hspace{1pt}U\hspace{1pt}|=1\}
$$
acting on $\mathbb{C}^2$.\footnote{\emph{Cf}.\ \citet{Weyl1929} p.~334: ``\foreignlanguage{german}{Das Transformationsgesetz der $\psi$-Komponenten besteht darin, daß sie unter dem Einfluß einer Transformation $\Lambda$ der Weltkoordinaten $x\textrm{(}\alpha\textrm{)}$ sich so umsetzen, daß die Größen [(\ref{QuadForm})] die Transformation $\Lambda$ erleiden.}''\label{erleiden}} The stronger condition $\mathrm{det}\hspace{1pt}U=1$ gives
$$
\mathbb{SL}\textrm{(}2,\mathbb{C}\textrm{)}=\{U\in\mathbb{GL}\textrm{(}2,\mathbb{C}\textrm{)}: \mathrm{det}\hspace{1pt}U=1\}.
$$
As the Lie algebras are the same, it makes \emph{local} sense to think of $\mathbb{W}$ as an extension\footnote{\emph{Cf}. \citet{Scholz2004} p.~189; \citet{Scholz2011b}, third page of the paper; and footnote \ref{extensions} above.}
$\mathbb{SL}\textrm{(}2,\mathbb{C}\textrm{)}\times\mathbb{U}\textrm{(}1\textrm{)}$ of $\mathbb{SL}\textrm{(}2,\mathbb{C}\textrm{)}$ by $\mathbb{U}\textrm{(}1\textrm{)}$. But multiplication of the `Cartesian factors' gives $(U,e^{i\theta})\mapsto e^{i\theta}U$, as well as $(-U,-e^{i\theta})\mapsto e^{i\theta}U$. Since $U\neq -U$ and $e^{i\theta}\neq -e^{i\theta}$, the pairs $(U,e^{i\theta})$ and $(-U,-e^{i\theta})$ must be \emph{distinct} elements of the extension $\mathbb{S}=\mathbb{SL}\textrm{(}2,\mathbb{C}\textrm{)}\times\mathbb{U}\textrm{(}1\textrm{).}$ So we have a 2-1 homomorphism
$$
f:\mathbb{S}\rightarrow\mathbb{W}\hspace{2pt}; (U,e^{i\theta})\mapsto e^{i\theta}U
$$
with inverse image $f^{-1}(e^{i\theta}U)=\{(U,e^{i\theta}),(-U,-e^{i\theta})\}$.\footnote{Here I am indebted to Thierry Levasseur and Johannes Huisman.}

There is also a 2-1 homomorphism\footnote{\citet{Weyl1929} p.~333: ``Man kann ihn normalisieren durch die Forderung, daß die Determinante von $U$ gleich $1$ sei, aber selbst dann bleibt eine Doppeldeutigkeit zurück.''}
$$
h: \mathbb{SL}\textrm{(}2,\mathbb{C}\textrm{)}\rightarrow \mathbb{SO^+}\textrm{(}1,3\textrm{)}\,;\hspace{2pt}U\mapsto \langle\sigma^\mu,U\sigma^\nu U^{\dagger}\rangle
$$
between $\mathbb{SL}\textrm{(}2,\mathbb{C}\textrm{)}$ and the component\footnote{\citet{Weyl1929} p.~333: ``\foreignlanguage{german}{wir unter den Lorentztransformationen nur die ein einziges in sich abgeschlossenes Kontinuum bildenden $\Lambda$ bekommen, welche 1.\ Vergangenheit und Zukunft nicht vertauschen und 2.\ die Determinante $+1$, nicht $-1$, besitzen [\,\dots].}''} $\mathbb{SO}^+\textrm{(}1,3\textrm{)}$ of $\mathbb{O}\textrm{(}1,3\textrm{)}$ containing the identity $\mathds{1}_4$; the inverse image
$$h^{-1}(\mathds{1}_4)=\{\pm\mathds{1}_2\}\subset\mathbb{SL}\textrm{(}2,\mathbb{C}\textrm{)}$$
contains $-\mathds{1}_2$ too---indeed $h\hspace{1pt}\textrm{(}U\textrm{)}=h\hspace{1pt}\textrm{(}\hspace{-2pt}-\hspace{-2pt}U\textrm{)}$ for all $U$. The matrix
$$
\left(\hspace{-4pt}
\begin{array}{cc}
 e^{i\omega} & 0  \\
 0 & e^{i\omega'}  \\
\end{array}
\hspace{-5pt}\right)\in\mathbb{U}\textrm{(}2\textrm{)}\subset\mathbb{W}
$$
belongs to $\mathbb{SU}\textrm{(}2\textrm{)}\subset\mathbb{SL}\textrm{(}2,\mathbb{C}\textrm{)}$ if $\omega' = -\omega$. The two eigenvalues would then have opposite phases, so the corresponding $\mathbb{O}\textrm{(}3\textrm{)}$ rotation (given by the quotient ${e^{i\omega'}}/{e^{i\omega}}$ of the eigenvalues, and hence by the difference of the phases) is twice the $\mathbb{SU}\textrm{(}2\textrm{)}$ phase angle $\omega$.\footnote{\citet{GTQM} p.129: ``\foreignlanguage{german}{Und zwar ist, wenn $\varepsilon=e^{i\omega}=e\textrm{(}\omega\textrm{)}$ gesetzt wird, der Drehwinkel der Drehung um die $z$-Achse $\varphi=-2\hspace{1pt}\omega$.}''} The homomorphism\footnote{I use the same letter $h$ for the restriction to $\mathbb{SU}\textrm{(}2\textrm{)}$.} $h:\mathbb{SU}\textrm{(}2\textrm{)}\rightarrow \mathbb{SO}\textrm{(}3\textrm{)}$ therefore maps \emph{iso}morphically on the half-open interval $0\leq\omega <\pi$; the isomorphism only breaks down at $\omega =\pi$, for there the negative identity $-\mathds{1}_2$ is reached, which already corresponds to the identity $\mathds{1}_4=h\hspace{1pt}\textrm{(}\hspace{-2pt}\pm\hspace{-2pt}\mathds{1}_2\textrm{)}$.\footnote{\citet{GTQM} p.~129: ``\foreignlanguage{german}{sie bleibt nämlich zweideutig, weil sie durch Multiplikation mit $-1$, durch Verwandlung von $\sigma$ in $-\sigma$ nicht verloren geht. [\,\dots] Man erhält dadurch alle Drehungen und jede genau zweimal.}''}
\[\begin{tikzcd}
\mathbb{U}\textrm{(}2\textrm{)}\arrow[hookrightarrow]{d} & \mathbb{SU}\textrm{(}2\textrm{)}\arrow[Rightarrow]{r}\arrow[hookrightarrow]{d}\arrow[hookrightarrow]{l} & \mathbb{SO}\textrm{(}3\textrm{)}\arrow[hookrightarrow]{d}\\
\mathbb{W} & \mathbb{SL}\textrm{(}2,\mathbb{C}\textrm{)}\arrow[Rightarrow]{r}\arrow[hookrightarrow]{l} & \mathbb{SO^+}\textrm{(}1,3\textrm{)}
\end{tikzcd}\]
The double arrows $\Rightarrow$ denote 2-1 homomorphisms, the hooked arrows $\hookrightarrow$ injections. The bottom row is relativistic and hyperbolic whereas the top row is compact, `spatial' and more literally `spherical.'

The orbit under $\mathbb{SL}\textrm{(}2,\mathbb{C}\textrm{)}$ of a timelike, future directed space-time four-vector satisfying
\begin{equation}\label{normal}
\mathrm{det}\hspace{1pt}\mathfrak{x}=\|\mathfrak{x}\|_{\eta}^2=1
\end{equation}
is the unit future hyperboloid $\mathfrak{H}_1^+$. If $\mathfrak{x}$ is written in spectral form $\mathfrak{x}=\zeta P+\zeta'P^{\perp}$, where the eigenvalues $\zeta$ and $\zeta'$ are real and $P$ is the projector $|\psi\rangle\langle\psi |$  along $|\psi\rangle\in\mathbb{C}^2$, normalization (\ref{normal}) is given by $\zeta'={1}/{\zeta}$. Cutting the hyperboloid by simultaneity surfaces $\Sigma_t$ orthogonal to $\sigma^0$ we obtain the spherical orbits\footnote{\citet{GTQM} p.~129: ``\foreignlanguage{german}{Jede \emph{unitäre} Transformation [\,\dots] liefert danach eine Drehung $s$ der Kugel [\,\dots]}.''} $\mathfrak{H}_1^+\cap\Sigma_t$ of $\mathbb{SU}\textrm{(}2\textrm{)}$, whose rotations affect only the eigen\emph{vectors} of $\mathfrak{x}$, not its eigenvalues. The spacelike ray $[r(\mathfrak{x}^1,\mathfrak{x}^2,\mathfrak{x}^3)]_r$ determines (the polar angles $\varphi$, $\theta$ of) the point on the sphere. The degenerate Hermitian operator $\mathfrak{x}=\mathds{1}_2$ (with $\zeta = \zeta' =1$) represents \emph{rest}, at the correspondingly degenerate bottom of the hyperboloid, where the sphere has shrunk to a point. Just as the orbits of $\mathbb{SU}\textrm{(}2\textrm{)}$ are horizontal, the \emph{pure boosts} making up the `complementary' part of $\mathbb{SL}\textrm{(}2,\mathbb{C}\textrm{)}$ displace vertically, straight up and down the hyperboloid, by affecting only the \emph{rest} of $\mathfrak{x}$: its eigen\emph{values}. A boost $B$ is `pure' with respect to the eigenvectors of $\mathfrak{x}$, which it therefore has to share, yielding an operator $B=\beta P+\beta^{-1}P^{\perp}$ of the same form, where $\beta$ is also real.\footnote{\citet{GTQM}, bottom of p.~131; a ``\foreignlanguage{german}{Zeitachse ändernde Lorentztransformation}'' is a boost.} Generic elements of $\mathbb{SL}\textrm{(}2,\mathbb{C}\textrm{)}$ which affect \emph{all} of $\mathfrak{x}$, eigenvalues and eigenvectors, produce `diagonal' motions on the hyperboloid. 

\subsubsection{Celestial sphere}
If $\zeta'$ vanishes we're left with $\mathfrak{x}=\zeta P$ or just the null ray\footnote{\citet{Weyl1929} p.~333: ``\foreignlanguage{german}{Die Variablen $\psi_1$, $\psi_2$ sowie die Koordinaten $x_\alpha$ kommen hier \so{nur ihrem Verh"{a}ltnis nach} in Frage.}''} containing $\psi$; null because $\mathrm{det}\hspace{1pt}\mathfrak{x}$ vanishes if $\zeta\zeta'$ does---the degenerate case $\zeta=\zeta'=0$ being at the origin, the tip of the cones. Rather than $\mathfrak{x}^{\mu}=\frac{1}{2}\mathrm{Tr}(P\sigma^{\mu})$ we can write $\mathfrak{x}^{\mu}=\langle\psi|\sigma^{\mu}|\psi\rangle$ or\footnote{\citet{Weyl1929} equations (2) p.~333 and (3) p.~334, \citet{GTQM} equations (8.12) and (8.16).}
\begin{equation}\label{QuadForm}
\mathfrak{x}^\mu=\bar{\psi}\sigma^\mu\psi.
\end{equation}
Since $\sigma^0$ is the identity $\mathds{1}_2$ and therefore $\mathfrak{x}^0$ equals $\langle\psi|\psi\rangle$, one may be tempted to write the Pythagorean expression $\langle\psi|\psi\rangle=\langle\psi|\sigma^1 |\psi\rangle+\langle\psi|\sigma^2 |\psi\rangle+\langle\psi|\sigma^3 |\psi\rangle$, which does not hold.\footnote{The terms look appropriately quadratic, the trinions $\sigma^k$ are indeed orthogonal but $\mathbb{R}^3$ should not be confused with $\mathbb{C}^2$.} Pythagorean equations that do hold are $(\mathfrak{x}^0)^2=(\mathfrak{x}^1)^2+(\mathfrak{x}^2)^2+(\mathfrak{x}^3)^2$ and $\langle\psi|\psi\rangle=|\psi_1|^2+|\psi_2|^2$, where $\psi_1=\langle\varphi_1|\psi\rangle$, $\psi_2=\langle\varphi_2|\psi\rangle$, and the basis $|\varphi_1\rangle,|\varphi_2\rangle$ is orthonormal. Now that we've gone from the hyperboloid to the future light cone $\mathfrak{K}^+$, the orbit of $\mathbb{SU}\textrm{(}2\textrm{)}$ is the \emph{celestial} sphere $S_t^+=\mathfrak{K}^+\cap\hspace{1pt}\Sigma_t$, which, choosing $t=1$, can be mapped to its equatorial plane $\mathbb{C}=\{(1,x,y,0)\}$ by stereographic projection from the south pole $(1,0,0,-1)$. I've written $\mathbb{C}$ because the coordinates $x,y$ of the equatorial plane can be viewed as the real and imaginary parts of the complex number $x+iy$, which is then identified with the quotient ${\psi_2}/{\psi_1}$ and hence the ray containing
$$
|\psi\rangle=\psi_1|\varphi_1\rangle+\psi_2|\varphi_2\rangle\textrm{,}
$$
where $|\varphi_2\rangle$ (or $\psi_1=0$) corresponds to the south pole. If $\psi_2$ vanishes, the ray from the south pole will be vertical, through the origin of $\mathbb{C}$ and the north pole, which therefore corresponds to $|\varphi_1\rangle$. Such a scheme works well for the `horizontal,' in other words `purely spatial' group $\mathbb{SU}\textrm{(}2\textrm{)}$; but boosts $\sigma^0\mapsto U\sigma^0 U^\dagger$ distort $S_t^+$ by tilting the surfaces $\Sigma_t$ perpendicular to $\sigma^0$. Rather than null vectors $\mathfrak{x}\in\mathfrak{K}^+$ we can take null rays $\rho=\textrm{[}r\mathfrak{x}\textrm{]}_r\subset\mathfrak{K}^+$. For a given foliation, the `space of (future) null rays' $\mathscr{S}^+$ is equivalent to a celestial sphere; but $\mathscr{S}^+$ behaves better under boosts by not relying on foliation.\footnote{See \citet{Weyl1929} p.~333.}

To the element $U\in\mathbb{W}$ which gives $\psi'=U\psi\in\mathbb{C}^2$ and hence $\mathfrak{x}'^\mu=\bar{\psi}' \sigma^\mu\psi'\in\mathbb{R}^4$ corresponds the element $\Lambda\in\mathbb{SO^+}\textrm{(}1,3\textrm{)}$ which returns $\mathfrak{x}^{\mu}=\Lambda^{\mu}_{\nu}\mathfrak{x}'^{\nu}$.\footnote{See footnote \ref{erleiden} above.} The quadratic form $\bar{\psi}\sigma^\mu\psi$ therefore establishes a correspondence between matter (in $\mathbb{C}^2$) and gravity (in $\mathbb{R}^4$). The loose angle\footnote{\citet{Weyl1929} p.~333: ``\foreignlanguage{german}{Durch $\Lambda$ ist die lineare Transformation $U$ der $\psi$ nicht eindeutig festgelegt, sondern es bleibt ein willkürlicher konstanter Faktor $e^{i\lambda}$ vom absoluten Betrage 1 zur Disposition. \emph{Cf}.\ \citet{GTQM} p.~131: ``\foreignlanguage{german}{Transformationen $\sigma$, welche sich nur durch einen Faktor $e^{i\lambda}$ vom absoluten Betrag $1$ voneinander unterscheiden, liefern dasselbe $s$.}''}\label{looseangle}} which will produce electricity below can already be seen in
$$\bar{\psi}e^{-i\lambda}\sigma^{\mu}e^{i\lambda}\psi=\bar{\psi}\sigma^\mu\psi.$$

\subsubsection{Electricity, gravity and matter\label{generation}}
From where, then, does electricity emerge?\footnote{See \citet{Weyl1929} p.~348, \citet{Rice} p.~291, \citet{WGA, EGM}.} Everything turns on the apparently insignificant choice\footnote{See footnote \ref{Wgroup} above.} of $\mathbb{W}$ over $\mathbb{SL}\textrm{(}2,\mathbb{C}\textrm{)}$, to propagate matter. One wonders how it can make any difference, for the two groups seem to differ by a mere nuance, but it is out of that nuance that Weyl extracts electricity.\footnote{\citet{Rice} p.~291: ``It is my firm conviction that we must seek the origin of the electromagnetic field in another direction. We have already mentioned that it is impossible to connect the transformations of the $\psi$ in a unique manner with the rotations of the axis system; however we may attempt to accomplish this by means of invariants which can be used as constituents of an action quantity we always find that there remains an arbitrary ``gauge factor'' $e^{i\lambda}$. Hence the local axis-system does not determine the components of $\psi$ uniquely, but only within such a factor of absolute magnitude 1.''} Various relevant quantities are invariant under $\mathbb{W}$, the best example\footnote{Here I am indebted to Ermenegildo Caccese.} being perhaps $\mathrm{det}(U\mathfrak{x}\hspace{1pt}U^{\dagger})$, with $U\in\mathbb{W}$; so the choice, however unusual, is by no means insensate. In a sentence: The laws governing matter and gravity are the same up to a detail that makes no difference to matter, but can nonetheless produce electricity.

The material connection $\mathfrak{M}$ is a one-form which, applied to a vector $\dot{\gamma}\in T_aM$ directed towards nearby $b\in M$, yields a generator $\langle \mathfrak{M},\dot{\gamma}\rangle$ of transport belonging to the Lie algebra $\mathfrak{w}=\mathrm{Lie}\hspace{1pt}\mathbb{W}$. The relationship between $\mathbb{W}$ and $\mathfrak{w}$ can be illustrated by looking at the subgroup $\mathbb{SU}\textrm{(}2\textrm{)}\subset\mathbb{W}$ and its Lie algebra $\mathfrak{su}\textrm{(}2\textrm{)}\subset\mathfrak{w}$. Taking
$$
U_{\tau}=e^{iE_1\tau}|\varphi_1\rangle\langle\varphi_1|+e^{iE_2\tau}|\varphi_2\rangle\langle\varphi_2|\in\mathbb{SU}\textrm{(}2\textrm{)}
$$
we have
$$
U_{\tau}|\psi\rangle=e^{i(E_1\tau+\eta_1)}|\psi_1||\varphi_1\rangle+e^{i(E_2\tau+\eta_2)}|\psi_2||\varphi_2\rangle\textrm{,}
$$
which represents a motion,\footnote{Notions of motion, speed and time are clearly metaphorical here, since $\tau$ is an abstract parameter.} infinitesimally generated by
$$
H=\langle \mathfrak{M},\dot{\gamma}\rangle=E_1|\varphi_1\rangle\langle\varphi_1|+E_2|\varphi_2\rangle\langle\varphi_2|\in\mathfrak{su}\textrm{(}2\textrm{),}
$$
on a two-dimensional torus determined by the two phases $\beta_k=E_k\tau+\eta_k$; the angles $\eta_k$ being the arguments of the coefficients $\psi_k$. Since $a=\gamma\textrm{(}0\textrm{)}$ and $b=\gamma\textrm{(}1\textrm{)}$ we have $\beta_{kb}=E_{ka}+\beta_{ka}$ and $e^{i\beta_{kb}}=e^{iE_{ka}}e^{i\beta_{ka}}$, and hence $|\psi_b\rangle=e^{iH}|\psi_a\rangle$. The generator $H$ acts only on the phase-torus and not on the moduli of $|\psi\rangle$; even infinitesimally, $|\psi\rangle$ has to be multiplied by a \emph{unitary} operator---multiplication by a Hermitian operator $H$ would lengthen the state by acting obliquely (rather than orthogonally). An element of the Lie algebra can be thought of here as an \emph{infinitesimal speed on the two-dimensional torus}. A group element produces an arbitrary rotation, the Lie algebra allows the single rotation to be broken down pointwise into its various infinitesimal rates. The distinction is somewhat obscured by a constant generator $H$, which produces a constant speed; a variable generator $H\textrm{(}\tau\textrm{)}$ with variable eigenvalues $E_k\textrm{(}\tau\textrm{)}$ would be needed to make full sense of the distinction. The geometrical picture of a motion on a torus suggests the conventional choice of viewing the generators of infinitesimal speed as \emph{real}, in other words Hermitian, rather than purely imaginary.

So the spinors representing matter are propagated by a connection with values in $\mathfrak{w}$; whereas gravity, represented by tetrads,\footnote{\citet{GTQM} p.~195: ``\foreignlanguage{german}{Ferner bedarf man in der allgemeinen Relativitätstheorie an jeder Weltstelle $P$ eines aus vier Grundvektoren in $P$ bestehenden normalen Achsenkreuzes, um die Metrik in $P$ festzulegen und relativ dazu die Wellengröße $\psi$ durch ihre vier [full Dirac theory, with mass] Komponenten $\psi_{\varrho}$ beschreiben zu können; die gleichberechtigten normalen Achsenkreuze in einem Punkte gehen durch die Lorentztransformationen auseinander hervor.}''} is governed by a connection with values in $\mathfrak{o}\textrm{(}1,3\textrm{)}.$ Weyl saw an apparently uninteresting Lie algebra $\mathfrak{w}\ominus\mathfrak{o}\textrm{(}1,3\textrm{)}$---caught, as it were, between matter and gravity---in which a third connection would surely take its values.\footnote{\citet{Weyl1929} p.~348: ``\foreignlanguage{german}{Dann ist aber auch die infinitesimale lineare Transformation $dE$ der $\psi$, welche der infinitesimalen Drehung $d\gamma$ entspricht, nicht vollständig festgelegt, sondern $dE$ kann um ein beliebiges rein imaginäres Multiplum $i\cdot df$ der Einheitsmatrix vermehrt werden}.'' \citet{Rice} p.~291: ``Then there remains in the infinitesimal linear transformation $dE$ of $\psi$, which corresponds to the given infinitesimal rotation of the axis-system, an arbitrary additive term $+id\varphi\cdot 1$.''} As in 1918, Weyl identified the real-valued connection with the electromagnetic potential\footnote{\citet{Rice} p.~291, \citet{GTQM} p.~195: ``\foreignlanguage{german}{Aus der Natur, dem Transformationsgesetz der Größe $\psi$ ergibt sich, da{\ss} die vier Komponenten $\psi_{\varrho}$ relativ zum lokalen Achsenkreuz nur bis auf einen gemeinsamen Proportionalitätsfaktor $e^{i\lambda}$ durch den physikalischen Zustand bestimmt sind, dessen Exponent $\lambda$ willkürlich vom Orte in Raum und Zeit abhängt, und daß infolgedessen zur eindeutigen Festlegung des kovarianten Differentials von $\psi$ eine Linearform $\sum_{\alpha}f_{\alpha}dx_{\alpha}$ erforderlich ist, die so mit dem Eichfaktor in $\psi$ gekoppelt ist, wie es das Prinzip der Eichinvarianz verlangt.}''} $A$, whose curvature $F=dA$ gave the electromagnetic field. Its derivative $dF=0$ in turn provided Maxwell's two homogeneous equations.\footnote{\citet{Weyl1929} p.~349, \citet{Rice} pp.~291-2. \emph{Cf}.\ \citet{Ryckman2009} p.~295: ``Weyl derived the Maxwell equations from the requirement of local phase invariance, thus coupling charged matter to the electromagnetic field, and so originating the modern understanding of the principle of local gauge invariance (``\emph{local symmetries dictate the form of the interaction}'') that lies at the basis of contemporary geometrical unification programs in fundamental physics.''} Infinitesimally, one can think of electricity as an appropriate `difference' $\ominus$ between matter and gravity (and integrally, as a quotient of sorts; but see \S\ref{anachronism}).

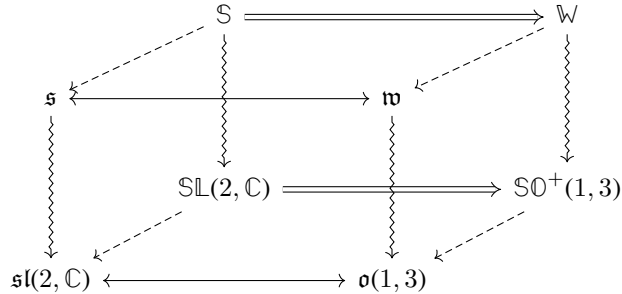
\[\begin{tikzcd}
& \mathbb{S} \arrow[dashrightarrow]{dl} \arrow[Rightarrow]{rr} \arrow[rightsquigarrow]{dd} & & \mathbb{W} \arrow[dashrightarrow]{dl} \arrow[rightsquigarrow]{dd}\\
\mathfrak{s} \arrow[leftrightarrow]{rr} \arrow[rightsquigarrow]{dd} & & \mathfrak{w}\\
& \mathbb{SL}\textrm{(}2,\mathbb{C}\textrm{)} \arrow[dashrightarrow]{dl} \arrow[Rightarrow]{rr} & & \mathbb{SO}^+\textrm{(}1,3\textrm{)} \arrow[dashrightarrow]{dl}\\
\mathfrak{sl}\textrm{(}2,\mathbb{C}\textrm{)} \arrow[leftrightarrow]{rr} & & \mathfrak{o}\textrm{(}1,3\textrm{)} \arrow[from=uu, rightsquigarrow]\\
\end{tikzcd}\]
As the `bivalency'\footnote{Weyl says \emph{Doppeldeutigkeit}, both homomorphisms are 2-1.}  of the homomorphisms represented by the double arrows $\Rightarrow$ is only global and not local, it is eliminated by the dashed arrows $\dashrightarrow$ (from the groups to their Lie algebras), which therefore give rise to the \emph{iso}morphisms denoted by the arrows $\leftrightarrow$ with two heads. All four projections---indicated by the vertical arrows---do away with a real number, which is an angle only for the `far' two, involving groups; an angle being a real number with a global structure to which differentiation (or local linearization) has no access. Since the real numbers eliminated by the two far projections (involving groups) are already present \emph{locally}, they survive the differentiation denoted by $\dashrightarrow$. An abuse of notation is worth pointing out: the dashed arrows map a group, \emph{treated as a single element}, to its Lie algebra (one sometimes writes $\mapsto$ between individuals), whereas the other arrows map, more conventionally, from the domain to the range.

The diagram can be extended by the projections
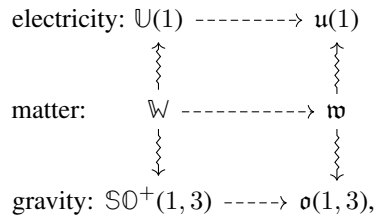
\[\begin{tikzcd}
\textrm{electricity:}\hspace{-41pt} & \mathbb{U}\textrm{(}1\textrm{)} \arrow[dashrightarrow]{r} & \mathfrak{u}\textrm{(}1\textrm{)}\\
\textrm{matter:}\hspace{-26pt} & \mathbb{W}  \arrow[dashrightarrow]{r} \arrow[rightsquigarrow]{d} \arrow[rightsquigarrow]{u} & \mathfrak{w} \arrow[rightsquigarrow]{d} \arrow[rightsquigarrow]{u} \\
\textrm{gravity:}\hspace{-29pt} & \mathbb{SO}^+\textrm{(}1,3\textrm{)} \arrow[dashrightarrow]{r} & \mathfrak{o}\textrm{(}1,3\textrm{),} \end{tikzcd}\]
which are complementary (as in ``orthogonal \emph{complement}'') in the sense that the arrow going down/up omits/keeps what's kept/omitted by the one going up/down.

The `angular freedom'\footnote{See footnote \ref{looseangle}.} caught between matter and gravity can be seen in
$$h'\textrm{(}U\textrm{)}=h'(e^{i\lambda}U)\in \mathbb{SO}^+\textrm{(}1,3\textrm{)}$$
or even
$$h'^{-1}(h'\textrm{(}U\textrm{)})=[e^{i\lambda}U]_{\lambda}\subset\mathbb{W}\textrm{,}$$
where $h':\mathbb{W}\rightarrow\mathbb{SO}^+\textrm{(}1,3\textrm{)}$ is given by the same rule $U\mapsto\langle\sigma^{\mu},U\sigma^{\nu}U^{\dagger}\rangle$, which also shows that the free angle $\lambda$ never reaches $\mathbb{SO}^+\textrm{(}1,3\textrm{)}$:
$$
e^{i\lambda}U\mapsto\langle\sigma^{\mu},e^{i\lambda}U\sigma^{\nu}e^{-i\lambda}U^{\dagger}\rangle=\langle\sigma^{\mu},U\sigma^{\nu}U^{\dagger}\rangle.
$$
Again, the fact that the `circle-1' homomorphism $h'$ maps $e^{i\lambda}U$ to a single Lorentz transformation for all angles $\lambda$ should not be confused with the fact that the 2-1 homomorphism $f:\mathbb{S}\rightarrow\mathbb{W}$ maps the two distinct pairs $(U,e^{i\theta})$ and $(-U,-e^{i\theta})$ to the same product $e^{i\theta}U\in\mathbb{W}$. Returning to the Lie algebras $\mathfrak{o}\textrm{(}1,3\textrm{)}$ and $\mathfrak{w}=\mathfrak{sl}\textrm{(}2,\mathbb{C}\textrm{)}\oplus \mathbb{R}\mathds{1}_2$, we have the corresponding expressions
$$\mathfrak{h}\textrm{(}\mathfrak{U}\textrm{)}
=\mathfrak{h}\textrm{(}\mathfrak{U}\oplus\lambda \mathds{1}_2\textrm{)}\in\mathfrak{o}\textrm{(}1,3\textrm{)}$$
and
$$\mathfrak{h}^{-1}(\mathfrak{h}\textrm{(}\mathfrak{U}\textrm{)})
=[\mathfrak{U}\oplus\lambda\mathds{1}_2]_{\lambda}\subset \mathfrak{w}\textrm{,}$$
where $\mathfrak{h}:\mathfrak{w}\rightarrow\mathfrak{o}\textrm{(}1,3\textrm{)}\hspace{1pt};$ $\mathfrak{U}\oplus\lambda \mathds{1}_2\mapsto\mathfrak{U}$ gets rid of $\lambda$ by projecting the pair $\textrm{(}\mathfrak{U},\lambda\textrm{)}$ to $\mathfrak{U}$. The loose phase $e^{i\lambda}\in\mathbb{U}\textrm{(}1\textrm{)}$ has become the `additive' freedom $\lambda\in\mathbb{R}=\mathfrak{u}\textrm{(}1\textrm{)}$.

So there's a connection for spinors, another for tetrads, and a third---namely $A$---for the residual $\mathbb{U}\textrm{(}1\textrm{)}$ freedom caught in between.\footnote{\citet{Weyl1929} p.~348: ``\foreignlanguage{german}{Zur eindeutigen Festlegung des kovarianten Differentials $\delta\psi$ von $\psi$ hat man also außer der Metrik in der Umgebung des Punktes $P$ auch ein solches $df$ für jedes von $P$ ausgehende Linienelement $\overrightarrow{PP}'=\textrm{(}dx\textrm{)}$ nötig}. Damit $\delta\psi$ nach wie vor linear von $dx$ abhängt, muß
$$
df=f_p\textrm{(}dx\textrm{)}^p
$$
eine Linearform in den Komponenten des Linienelements sein. Ersetzt man $\psi$ durch $e^{i\lambda}\cdot\psi$, so muß man sogleich, wie aus der Formel für das kovariante Differential hervorgeht, $df$ ersetzen durch $df-d\lambda$.'' \citet{Rice} p.~291: ``The complete determination of the covariant differential $\delta\psi$ of $\psi$ requires that such a $d\varphi$ be given. But it must depend linearly on the displacement $PP'$: $d\varphi=\varphi_p\textrm{(}dx\textrm{)}^p$, if $\delta\psi$ shall depend linearly on the displacement. On altering $\psi$ by multiplying it by the gauge factor $e^{i\lambda}$ we must at the same time replace $d\varphi$ by $d\varphi-d\lambda$ as is immediately seen from this formula of the covariant differential.'' Weyl's notation is confusing: whereas the one-form $d\lambda$ (which \emph{is} a differential) is necessarily exact, $df$ and $d\varphi$ (my $A$) aren't.} The values $\langle A,\dot{\gamma}\rangle$ belong to the Lie algebra $\mathfrak{u}\textrm{(}1\textrm{)}$ of the group $\mathbb{U}\textrm{(}1\textrm{)}$ caught between matter and gravity. The gravitational connection
$$
\mathit{\Gamma}=\mathit{\Gamma}_{\mu}\otimes dx^\mu=\mathit{\Gamma}^r_\mu\mathbf{T}_r\otimes dx^\mu
$$
takes its values $\langle \mathit{\Gamma},\dot{\gamma}\rangle=\mathit{\Gamma}^r_\mu\dot{\gamma}^\mu \mathbf{T}_r$ in $\mathfrak{o}\textrm{(}1,3\textrm{)}$, the material connection
$$
\mathfrak{M}=\mathfrak{M}_\mu\otimes dx^\mu=\mathfrak{M}^r_\mu \mathbf{T}_r\otimes dx^\mu
$$
its values $\langle \mathfrak{M},\dot{\gamma}\rangle=\mathfrak{M}^r_\mu\dot{\gamma}^\mu \mathbf{T}_r$ in $\mathfrak{w}$. The three connections are related by their Lie algebras
$$\mathrm{Lie}\,\mathbb{W}=\mathrm{Lie}\,\mathbb{SL}\textrm{(}2,\mathbb{C}\textrm{)}\oplus \mathrm{Lie}\,\mathbb{U}\textrm{(}1\textrm{).}$$
For an electron subject to gravity as well as electricity we can write
\begin{equation}\label{CovWE}
\slashed D\psi=0
\end{equation}
instead of (\ref{WEquation}), where $\slashed D=\sigma^{\mu}D_\mu$ and\footnote{This doubly covariant derivative for matter interacting with electricity \emph{and} gravity is obtained by combining the `gravitational covariance' expressed in equation (13) of \citet{Weyl1929} with the `electromagnetic covariance' expressed at the bottom of p.~350 and especially the top of p.~351, same paper.}
$$
\partial_\mu\mapsto D_{\mu}=\partial_\mu +i\mathfrak{M}_{\mu}=\partial_\mu +i(\mathit{\Gamma}_{\mu}+A_{\mu}).
$$
Already in \textsf{\small{W18}} one could express the total---rotating and dilating---connection as an appropriate sum $\mathit{\Gamma}+A$ of a metric (purely gravitational) connection $\mathit{\Gamma}$ and a dilating (electric) connection $A$; adding electricity to gravity one obtained the two together, not something new, a third element. In \textsf{\small{W29}} we have the same two terms, but they add up to \emph{matter}.

This account does not render all the historical colour of Weyl's argument,\footnote{\citet{Weyl1929} p.~348, \citet{Rice} p.~291} which I can try to express more faithfully as follows. If the phase angle $\lambda$ were propagated holonomically by $A$, the curvature  $F=dA$ would vanish and electricity with it. To convince himself that $\lambda$ has to vary anholonomically, Weyl relates its propagation to that of the tetrads representing gravitation. He seems to argue that if the tetrad were `constant,' $\lambda$ would be too; since the tetrad varies, $\lambda$ should too. \emph{Holonomy} is the best meaning I can give to the \emph{constancy} of a tetrad. Only a \emph{flat} gravitational connection $\mathit{\Gamma}$ allows the assignment of the \emph{same} tetrad to different points; only with flatness can there be \emph{global} constancy or `sameness'; with curvature it becomes meaningless to say that tetrads at different points are the same. Where the constancy of tetrads makes no sense, one can suppose they \emph{vary}. Since the tetrad's variation is given by infinitesimal propagation, so is $\lambda$'s; in any case there is no reason to confine $\lambda$'s variation (holonomically) to a continuous function $\lambda: M\rightarrow\mathbb{R}$, which would be too restrictive. The object needed for the infinitesimal propagation of an angle, linearly in the angle and the direction of propagation $\dot{\gamma}$, is a real-valued one-form $A$. A few words about a possible confusion: With a non-Abelian Lie algebra $\mathfrak{g}$ acting on a (nontrivial) vector space $\mathbb{V}$ (such as $\mathbb{C}^N$ with $N>1$), there can be no confusion between the operator $\langle\mathfrak{M},\dot{\gamma}\rangle\in\mathfrak{g}$ and its argument in $\mathbb{V}$, which are mathematical objects of different kinds; the matrix representations are of different shapes and sizes, $N\times N$ rather than $N$. But where $\langle A,\dot{\gamma}\rangle$ is a real number acting on another real number $\lambda$, the operator and its argument are easily confused. Here the angle is acted upon by a `scalar' operator $\langle A,\dot{\gamma}\rangle\in\mathfrak{u}\textrm{(}1\textrm{)}$.

Weyl claims that in special relativity there's a single tetrad and hence just one value of $\lambda$,\footnote{\citet{Weyl1929} p.~348: ``\foreignlanguage{german}{In der speziellen Relativitätstheorie muß man diesen Eichfaktor als eine Konstante ansehen, weil wir hier ein einziges, nicht an einen Punkt gebundes Achsenkreuz haben}.'' \citet{Rice} p.~291: ``In the special theory of relativity, in which the axis system is not tied up to any particular point, this factor is a constant."} which may mean that the structure groups $G=\mathbb{SO}^+\textrm{(}1,3\textrm{)}$ and $G'=\mathbb{R}$ (acting at a generic space-time point) coincide with the corresponding gauge groups\footnote{A gauge group is made up of \emph{vertical automorphisms} $\upsilon:E\rightarrow E$ on the (here trivial) fibre bundle $E=M\times\mathbb{V}$---\emph{vertical} inasmuch as each copy $G_x$ of $G$ confines its action to its own $\mathbb{V}_x$, without interfering with the other fibres $\mathbb{V}_{x'}$. Since \emph{horizontal} is a metaphor for `constancy' from fibre to fibre along $M$, \emph{vertical} means `just up the fibre' (and not along $M$). However `symmetric' the Cartesian product $\cdot\times\cdot$  may look, here it isn't at all, since the two factors are distinguished as \emph{base} $M$ and \emph{fiber} $\mathbb{V}$: a copy $\mathbb{V}_x$ gets assigned to each $x$ of the base manifold so that $x$ can be fixed while $\psi\in\mathbb{V}_x$ is varied, whereas `displacement only along the base manifold with constancy in the corresponding fibers $\{\mathbb{V}_x\}_x$' makes no sense without further structure, namely a connection.} $\mathscr{G}\simeq G$ and\footnote{This equivalence with the structure group expresses the `constant' degeneracy of the gauge group, which, having lost all the pointwise freedom to vary its action over the underlying manifold, rigidly applies the same element $\mathrm{g}\in G$ everywhere.} $\mathscr{G}'\simeq G'$ (acting---rigidly here---on all of space-time $M$).\footnote{\emph{Cf}.\ \citet{Kretschmann}: \emph{general} covariance can be countenanced in \emph{flat} space-time.} In general relativity\footnote{\citet{Weyl1929} p.~348: ``\foreignlanguage{german}{Anders in der allgemeinen Relativitätstheorie: jeder Punkt hat sein eigenes Achsenkreuz und darum auch seinen eigenen willkürlichen Eichfaktor; dadurch, daß man die starre Bindung der Achsenkreuze in verschiedenen Punkten aufhebt, wird der Eichfaktor notwendig zu einer willkürlichen Ortsfunktion}.'' \citet{Rice} p.~291: ``But it is otherwise in the general theory of relativity when we remove the restriction binding the local axis-systems to each other; we cannot avoid allowing the gauge factor to depend arbitrarily on position.''} the tetrad can vary, and $\lambda$ too; the gauge groups $\mathscr{G}$ and $\mathscr{G}'$, which no longer act rigidly,\footnote{\emph{Cf}.\ \citet{Weyl1929} p.~331: ``\foreignlanguage{german}{es fällt mir schwer, die Macht zu begreifen, welche die lokalen Achsenkreuze in den verschiedenen Weltpunkten in ihrer verdrehten Lage zu starrer Gebundenheit aneinander hat einfrieren lassen.}''} become much (indeed infinitely) larger than the structure groups.\footnote{Of course the flatness of space-time only imposes \emph{holonomy} $\mathscr{G}_H\subset\mathscr{G}$, not \emph{rigidity} $G\subset\mathscr{G}_H$, which is much stronger; \emph{cf}.\ \citet{Ryckman2009} p.~295: ``Weyl's argument for his correct conclusion is, in fact, flawed, resting on an unnecessary assumption about the representation of spinor matter fields within tetrad formulations of arbitrarily curved space-times.'' The ``flatness'' I mean refers to the Levi-Civita connection, which is both metric and symmetric; a metric connection with torsion can produce (torsional) anholonomies, even on Minkowski space-time.}

But why, one may ask, should the variations of $\lambda$ and tetrads be at all related in the first place? Are they not independent? Why not a curved $A$ with a flat $\mathit{\Gamma}$, or the other way around? We may simply have another case of \emph{geometrical justice}: since tetrads are allowed to vary anholonomically, why not $\lambda$ too?\footnote{\emph{Cf}.\ \citet{Weyl1929} pp.~331-2: ``\foreignlanguage{german}{Gerade dadurch, daß man den Zusammenhang zwischen den lokalen Achsenkreuzen löst, verwandelt sich der Eichfaktor $e^{i\lambda}$, der in der Größe $\psi$ willkürlich bleibt, notwendig aus einer Konstante in eine willkürliche Ortsfunktion; d.\ h. nur durch diese Lockerung wird die tatsächlich bestehende Eichinvarianz verständlich.}''} Even if the length and direction that deserved the same treatment in 1918 were part of a single object,\footnote{What does or doesn't constitute a `single object' is rather arbitrary: the direction and length of a vector can be brought apart by taking, instead of a vector, a ray (one object) and a separate number (another object); there are likewise ways of building a single object out of a number and a tetrad.} they could vary just as independently as $\lambda$ and tetrads: a flat length connection with a curved directional connection, or even the other way around (which gives electricity without gravity---in flat space-time).

In 1918 all Weyl had to identify electricity was the electromagnetic look of the expressions $F=dA$ and $dF=0$; but the Hamiltonian, quantum-mechanical content of \textsf{\small{W29}} provides more. Electricity is represented in Hamiltonian theory by adding the electromagnetic potential to momentum:\footnote{See \citet{GTQM} p.~88.}
\begin{equation}\label{CMomentum}
p\mapsto p+A
\end{equation}
or $p_\mu\mapsto p_\mu+A_\mu$. Again, momentum in quantum mechanics is given by differentiation:
\begin{equation}\label{QMomentum}
p\mapsto\hat{p}= -id
\end{equation}
or $p_\mu\mapsto\hat{p}_\mu= -i\partial_\mu$. The rule\footnote{See \citet{Rice} p.~283, \citet{GTQM} p.~89.}
\begin{equation}\label{covariant}
d\mapsto D=d+iA
\end{equation}
or $D_{\mu}=\partial_{\mu}+iA_{\mu}$
is obtained by combining (\ref{CMomentum}) and (\ref{QMomentum}). The compensation of (\ref{phase}) by (\ref{gauge}) can be seen in the Lagrangian
$$\mathscr{L}=\bar{\psi}\hspace{1pt}\sigma^\mu D_\mu\psi=\bar{\psi}'\sigma^\mu (D_\mu-i\partial_\mu\lambda)\psi'.$$
Here electricity is introduced via (\ref{CMomentum}), not conjured up; but once we have a real-valued connection $A$ (caught between matter and gravity), whose first two derivatives look electromagnetic, (\ref{CMomentum})-(\ref{covariant}) provide valuable support. Weyl nonetheless claims to have derived electricity \emph{independently} (without (\ref{CMomentum})), out of the group-theoretical relationship between matter and gravity.

\section{Yang-Mills\label{YangMills}}
\noindent Here the structure group $\mathbb{SU}\textrm{(}N\textrm{)}$ replaces $\mathbb{U}\textrm{(}1\textrm{)}$. Weyl is no longer in the foreground, nor is his complaint that Dirac's theory gave \textsf{\small{EL}}: \emph{twice too many energy levels}. The curved space-time from which \textsf{\small{W18}} arose can now, having done its bit,\footnote{The \emph{geometrical justice} of \S\ref{GeomJust} required a curved length connection $A$ to balance the curved directional connection. By adopting a flat space-time connection alongside a curved isospin connection \citet{YangMills} reversed the injustice of Einstein's theory---which has a curved directional connection with a flat length connection.} be kept or dropped.

Let us go back to (\ref{phase}), which is indeed a natural choice to replace (\ref{conforme}). But is it the \emph{only} natural choice? The transformation on the infinite-dimensional Hilbert space $\mathscr{H}$ containing $\psi$ should no doubt be unitary, but (\ref{phase}) is a very special unitary transformation, which ought perhaps to be generalized.

One thinks of the function $\lambda$ as `real-valued': it assigns a real number $\lambda_x$ to every $x\in M$. Since the wavefunction $\psi$ assigns not a complex number but a spinor $\psi_x\in\mathbb{C}^N_x$ to every $x$, the value $\lambda_x$ is in fact the operator
$$
\lambda_x\cdot\mathds{1}_N:\mathbb{C}^N_x \rightarrow\mathbb{C}^N_x.
$$
But then why not take a \emph{general} Hermitian operator $\Lambda_x:\mathbb{C}^N_x\rightarrow\mathbb{C}^N_x$ rather than the very special Hermitian operator $\lambda_x\cdot\mathds{1}_N$? \emph{Why stop when one's almost there?} Legitimate question, which is enough to yield \textsf{\small{YM}}. The infinitesimal generators $\Lambda$ give the Lie algebra $\mathfrak{su}\textrm{(}N\textrm{)}$ of the structure group $\mathbb{SU}\textrm{(}N\textrm{)}$, whose elements $e^{i\Lambda}$ act at a single space-time point. Again, the gauge group $\mathscr{G}$ acting on all of $M$ will not in general be a rigid copy of the structure group. The pointwise freedom to perform independent rotations at every $x\in M$ is obtained by putting together independent copies $\mathbb{SU}_x\textrm{(}N\textrm{)}$ of $\mathbb{SU}\textrm{(}N\textrm{)}$. The gauge group acts unitarily on $\mathscr{H}$ since every $e^{i\Lambda_x}\in\mathbb{SU}_x\textrm{(}N\textrm{)}$, point by point, acts unitarily on its fibre $\mathbb{C}^N_x$. One can think of the \emph{vertical} character of the automorphism $\mathscr{U}\in\mathscr{G}$ in (appropriately continuous) `block diagonal' terms, where the block $e^{i\Lambda_x}$ is identified by $x$.

Mathematically, the Yang-Mills connection $\mathscr{A}=\mathscr{A}_{\mu}^k \mathbf{U}_k\otimes dx^{\mu}$ can be seen as a unitary (and $N$-dimensional) version of Weyl's material connection $\mathfrak{M}$, in the sense that $\mathbb{SU}\textrm{(}N\textrm{)}$ replaces $\mathbb{W}$. The infinitesimal generators $\mathbf{U}_1,\dots,\mathbf{U}_N$ span the Lie algebra $\mathfrak{su}\textrm{(}N\textrm{)}$. Applied to a tangent vector $\dot{\gamma}\in T_aM$ directed towards nearby $b\in M$, the one-form $\mathscr{A}$ gives the generator
$$\langle \mathscr{A},\dot{\gamma}\rangle=\mathscr{A}_\mu \dot{\gamma}^{\mu}
=\mathscr{A}^k_{\mu}\langle dx^{\mu},\dot{\gamma}\rangle\mathbf{U}_k.$$

In 1929 Weyl would have seen no \emph{physical} reason to take the step from $\mathbb{U}\textrm{(}1\textrm{)}$ to $\mathbb{SU}\textrm{(}N\textrm{)}$. Though given to physicomathematical speculation of great imaginative virtuosity, he didn't take the purely mathematical step either. The details of the physics that ultimately did produce the non-Abelian theory are in \citet{YangMills}. So one can distinguish between the `purely logical' step (which I've called \textsf{\small{NA}}: \emph{non-Abelian structure group}) and the `physical' step that would be taken in the fifties. The logical step, however conceptual or even fictitious, seems just as relevant here.

\section{Logical summary}
Summing up, \textsf{\small{W18}} was given by \emph{geometrical justice} \textsf{\small{GJ}} applied to \textsf{\small{GR}}:
$$
\textsf{\small{GR \& GJ}} \Rrightarrow \textsf{\small{W18}}.
$$
To reach \textsf{\small{W29}}, \emph{matter wave} \textsf{\small{MW}}, \emph{second clock effect} \textsf{\small{SC}} and \emph{twice too many energy levels} \textsf{\small{EL}} were needed too:
$$
\textsf{\small{W18 \& MW \& SC \& EL}} \Rrightarrow \textsf{\small{W29}}.
$$
To obtain \textsf{\small{YM}} from \textsf{\small{W29}}, a \emph{non-Abelian structure group} \textsf{\small{NA}} would have been enough:
$$
\textsf{\small{W29 \& NA}} \Rrightarrow \textsf{\small{YM}}\textrm{,}
$$
or more precisely
$$
\textsf{\small{W18 \& MW \& SC \& NA}} \Rrightarrow \textsf{\small{YM}}.
$$

Weyl had the greatest creative freedom in 1918, when he applied \emph{geometrical justice} to \textsf{\small{GR}}. The next moves were more constrained. In introducing a \emph{matter wave} after the discoveries of Schrödinger \emph{et al}.\ he had little choice; and it had to be relativistic, hence with spin, which led to the use of tetrads. Weyl's preference for (\ref{gauge})\&(\ref{phase}) over (\ref{gauge})\&(\ref{conforme}) was dictated by Einstein's objection, the \emph{second clock effect}. His reaction to the \emph{too many energy levels} was less constrained, but also less right, less consequential, more idiosyncratic. The adoption of a \emph{non-Abelian structure group} was mathematically so natural as to be almost inevitable; but ultimately the step was not taken for purely mathematical reasons, and as a physical move it seems more creative, less predictable.

Articles are sometimes written to settle priority disputes. Even if I'm aware of none here---my purpose has been entirely different---I'll end with a few words that may have to do with a kind of `priority': Since the generalization from $1$ to $N$ in $\mathbb{SU}\textrm{(}N\textrm{)}$ is unremarkable \emph{in itself} (quite apart from any implications it may have), one possible interpretation of the transition $\textsf{\small{W29 \& NA}}\Rrightarrow\textsf{\small{YM}}$ is: \emph{it was \emph{almost} all there in 1929}. Of course the mathematical \emph{consequences} of $\mathbb{U}\textrm{(}1\textrm{)}\rightarrowtail\mathbb{SU}\textrm{(}N\textrm{)}$ are hardly trivial, and have to be spelled out, which takes doing. But however hard to work out, those consequences \emph{are} $\textsf{\small{YM}}$.

\vspace{12pt}
\noindent I thank Vieri Benci, Julien Bernard, Alexander Blum, Ermenegildo Caccese, Adam Caulton, Radin Dardashti, Jacopo Gandini, Johannes Huisman, Marc Lachièze-Rey, Thierry Levasseur, Jean-Philippe Nicolas, Roger Penrose, Thomas Ryckman, George Sparling and Karim Thébault for valuable conversations and corrections.

\end{document}